\documentclass[a4paper,11pt]{article}
\pdfoutput=1
\usepackage[utf8]{inputenc}
\usepackage{jheppub}
\usepackage{comment}
\usepackage{bm}
\usepackage[dvipsnames]{xcolor}
\usepackage{tikz}
\usetikzlibrary{positioning}
\usetikzlibrary{shapes.geometric}
\usetikzlibrary{ decorations.markings}
\usetikzlibrary{graphs,graphs.standard,quotes}
\usepackage{ifthen}
\usepackage {empheq}
\newcommand{\nicebox}{\fboxsep=8pt\fbox}

\newcommand{\D}{\mathrm{d}}

\renewcommand{\O}{\mathcal{O}}
\newcommand{\x}{\bs{x}}

\newcommand{\<}{\langle}
\renewcommand{\>}{\rangle}
\newcommand{\lla}{\langle \! \langle}
\newcommand{\rra}{\rangle \! \rangle}

\newcommand{\bs}[1]{\boldsymbol{#1}}

\newcommand{\nn}{\nonumber}

\renewcommand{\b}{\beta}

\newcommand{\ep}{\epsilon}

\renewcommand{\j}{\varphi}

\renewcommand{\[}{\begin{equation}}
\renewcommand{\]}{\end{equation}}

\begin{document}

\title{Shift operators from the simplex representation in momentum-space CFT}

\author{Francesca Caloro and Paul McFadden}
\affiliation{School of Mathematics, 
	Statistics \& Physics, Newcastle University, 
	Newcastle NE1 7RU, U.K.}

\emailAdd{f.caloro2@ncl.ac.uk, paul.l.mcfadden@ncl.ac.uk}

\abstract{ We derive parametric 
integral representations for the general $n$-point function of scalar operators in momentum-space conformal field theory.  Recently, this was shown to be expressible as a generalised Feynman integral with the topology of an $(n-1)$-simplex, featuring an arbitrary function of momentum-space cross ratios.  Here, we show all graph polynomials for this integral can be expressed in terms of the first and second minors of the  Laplacian matrix for the simplex.
Computing the effective resistance between nodes of the corresponding electrical network, an inverse parametrisation is found in terms of the determinant and first minors of the Cayley-Menger matrix. 
These parametrisations reveal new families of weight-shifting operators, expressible as determinants, that connect $n$-point functions
in spacetime dimensions differing by two.  
Moreover, the action of all previously known weight-shifting operators  preserving the spacetime dimension is manifest.  
Finally, the new parametric representations enable the validity of the conformal Ward identities to be established directly, without recourse to recursion in the number of points.

}

\maketitle

\section{Introduction}

Understanding 
the general form of correlation functions in momentum-space conformal field theory is an important goal.  
Working in momentum space is natural for many applications, particularly 
 inflationary cosmology (see, {\it e.g.,} \cite{Antoniadis:2011ib,Maldacena:2011nz, Creminelli:2011mw, Bzowski:2012ih, Mata:2012bx,Kehagias:2012pd, McFadden:2013ria,Ghosh:2014kba, Anninos:2014lwa,Arkani-Hamed:2015bza, Arkani-Hamed:2018kmz, Baumann:2019oyu,Baumann:2020dch,Sleight:2019hfp}), 
 and reveals  features inherited from scattering amplitudes that would otherwise be hidden, for example double-copy structure and colour/kinematics duality \cite{Raju:2012zr, Farrow:2018yni,Lipstein:2019mpu, Armstrong:2020woi,Albayrak:2020fyp}.  
Momentum-space methods are moreover well suited for renormalisation \cite{Bzowski:2015pba,Bzowski:2017poo,Bzowski:2018fql,Bzowski:2022rlz}, and are of growing interest
 for the conformal 
 bootstrap 
 \cite{Gillioz:2018mto, Gillioz:2019iye,Gillioz:2020wgw}. 
 
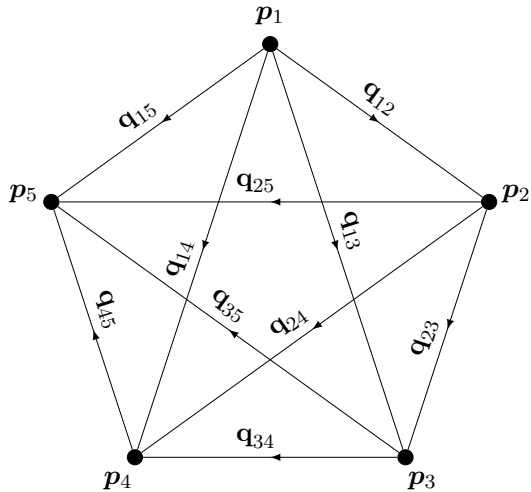
\begin{figure}[t]
\centering
\resizebox{0.55\textwidth}{!}{
\def\r{6pt}
\tikzset{d/.style={draw,circle,fill=black,minimum size=\r,inner sep=0pt, anchor=center}}

\begin{tikzpicture}[decoration={markings, 
    mark= at position 0.5 with {\arrow{latex}},
    }]

\pgfmathtruncatemacro{\Ncorners}{5}
\node[draw=white, regular polygon,regular polygon sides=\Ncorners,minimum size=8cm] 
(poly\Ncorners) {};
\node[regular polygon,regular polygon sides=\Ncorners,minimum size=8.7cm] 
(outerpoly\Ncorners) {};
\foreach\x in {1,...,\Ncorners}{
    \pgfmathtruncatemacro{\y}{90-(\x-1)*360/\Ncorners}
    \node[d] (poly\Ncorners-\x) at (\y:3.2){};
    \node (outerpoly\Ncorners-\x) at (\y:3.6){$\bs{p}_\x$};
}

\foreach \x in {1,...,5} {
  \foreach \y in {\x,...,5} {
    \ifthenelse{\x=\y}{}{\draw [postaction={decorate}](poly\Ncorners-\x)--node[auto,sloped]{$\hspace{-4mm}\textbf{q}_{\x\y}$}(poly\Ncorners-\y);}
}}

\end{tikzpicture}
}
\caption{Structure of the simplex integral, illustrated for the  $5$-point function.   \label{simplexfig}}
\end{figure}

In position space, the structure of general scalar $n$-point  functions has been understood for over fifty years \cite{Polyakov:1970xd}. 
A correspondingly general solution in momentum space was  proposed only
 recently in \cite{Bzowski:2019kwd, Bzowski:2020kfw}.   
This takes the form of a generalised 
Feynman integral with the topology of an $(n-1)$-simplex,
\begin{align} \label{simplex}
&\< \O_1(\bs{p}_1) \ldots \O_n(\bs{p}_n) \> =
 \prod_{1 \leq i < j \leq n} \int \frac{\D^d \bs{q}_{ij}}{(2 \pi)^d} \frac{f(\hat{\bs{q}})}{q_{ij}^{2 \alpha_{ij} + d}}  \prod_{k=1}^n (2\pi)^d \delta \Big( \bs{p}_k + \sum_{l=1}^n \bs{q}_{lk} \Big),
\end{align}
where  the integration is taken over the internal momenta $\bs{q}_{ij}$ running between vertices of the simplex.  Here $\bs{q}_{ij}=-\bs{q}_{ji}$ runs from vertex $i$ to $j$, while the external momenta  $\bs{p}_i$ enter only via momentum conservation as imposed by the delta function inserted at each vertex.    Each propagator corresponds to an edge of the simplex, as illustrated in figure \ref{simplexfig}, and is raised to a power specified by the parameter $\alpha_{ij}$.  Together, these satisfy the constraints
\begin{equation} \label{alphaijdef}
\Delta_i = - \sum_{j=1}^n \alpha_{ij}, 
\end{equation}
where $\Delta_i$ is the scaling dimension of the  operator $\mathcal{O}_i$.  To simplify the writing of such sums we define $\alpha_{ii} = 0$ and $\alpha_{ji} = \alpha_{ij}$. 
Euclidean signature will be assumed throughout.

The distinguishing  
feature of the simplex representation 
\eqref{simplex} is the presence of an {\it arbitrary function} $f(\hat{\bs{q}})$  of the independent 
momentum-space cross ratios  
\[ \label{conf_ratio_q}
\hat{q}_{[ijkl]} = \frac{q_{ij}^2 q_{kl}^2}{q_{ik}^2 q_{jl}^2},
\]
denoted collectively  by the vector $\hat{\bs{q}}$. 
As the simplex representation can be derived  from the general position-space solution \cite{Bzowski:2019kwd, Bzowski:2020kfw}, the number of independent cross ratios is the same in both cases, {\it i.e.,} $n(n-3)/2$ for $n\le d+2$ and $nd - (d+2)(d+1)/2$  for $n>d+2$. 
For $n\ge 4$, the solution of the constraints \eqref{alphaijdef} for the $\alpha_{ij}$ is not unique, but making a different choice simply multiplies $f(\hat{\bs{q}})$ by a product of powers of the cross ratios \eqref{conf_ratio_q}.  Since $f(\hat{\bs{q}})$ is arbitrary,  the solution of \eqref{alphaijdef}  chosen is therefore immaterial.

In this paper, we explore scalar parametric representations of the simplex integral \eqref{simplex} obtained by integrating out the internal momenta.  This offers several advantages:

\begin{itemize}
\item The original integral  \eqref{simplex} 
features $n(n-1)/2$ $d$-dimensional loop integrations and we have $(n-1)$ delta functions to help us, with one remaining behind to enforce overall momentum conservation.  This leaves the equivalent of $(n-1)(n-2)d/2$ scalar integrals to perform. 
In contrast, the  parametrisations we derive feature fewer integrals: only $n(n-1)/2$ scalar parametric integrals,
one for each edge of the simplex.

\item By inverting the graph  polynomials that arise, 
we construct novel weight-shifting operators connecting  solutions of the conformal Ward identities in spacetime dimension $d$ to new solutions in dimension $d+2$.  Remarkably, these operators have a determinantal structure 
based on the Cayley-Menger matrix familiar from 
distance geometry.  
In contrast, the well-known  weight-shifting operators 
 introduced in \cite{Karateev:2017jgd} preserve the spacetime dimension.  
Operators mapping  $d\rightarrow d+2$ are we believe 
known only for  $3$-point functions, where their existence can be seen from the  triple-$K$ representation in momentum space \cite{Bzowski:2015yxv}\footnote{These $d\rightarrow d+2$ operators also enable the construction of  $d$-dimensional {\it tensorial} correlators \cite{Bzowski:2013sza, Bzowski:2017poo,Bzowski:2018fql}.},
and for $4$-point  conformal blocks in position space  
(the operator $\mathcal{E}_+$ in \cite{Dolan:2011dv}).
The new $d\rightarrow d+2$ operators we obtain 
can be viewed as a natural generalisation of the $3$-point operators of \cite{Bzowski:2015yxv} 
to  arbitrary $n$-point correlators.
\end{itemize}

The plan of this paper is as follows.  In section \ref{sec1}, we show that all  graph polynomials for the simplex integral \eqref{simplex} can be constructed from the corresponding Gram matrix.
The standard parametric representations for Feynman integrals then follow. 
Alternatively, by regarding the  Schwinger parameters as resistances in an electrical network, we can  compute  the {\it effective} resistances between all vertices of the simplex.  This latter set of variables dramatically simplifies the structure of the Schwinger exponential.  In section \ref{sec:wsops}, we use these effective resistances to construct new $d\rightarrow d+2$ shift operators for the general $n$-point function.  The cases  $n=3,4$  are discussed in detail, and we verify 
the action of all operators independently 
through computation of their intertwining relations with the conformal Ward identities.  
The actions of
the   $d$-preserving weight-shifting operators of \cite{Karateev:2017jgd} are also demonstrated from this scalar parametric perspective. 
In section \ref{sec:CWI}, we  prove that the new parametric representations 
indeed solve the conformal Ward identities.  In contrast to the vectorial representation \eqref{simplex} (for which the Ward identities  are analysed in \cite{Bzowski:2019kwd, Bzowski:2020kfw}), 
for  the new scalar parametric representations the Ward identities can be verified directly without use of recursive arguments in the number of points $n$. 
As we show in section \ref{sec:posnspace},
the validity of the conformal  Ward identities, as well as the action of the $d$-preserving weight-shifting operators, can also be seen from  the position-space counterpart of the simplex. Section \ref{sec:disc} concludes with a summary of results and open directions.

\section{Parametric representations of the simplex}
\label{sec1}

This section investigates scalar parametric representations for the simplex integral \eqref{simplex}.
In the following, we identify the necessary graph polynomials (section \ref{sec:graphpolys}), standard parametric representations (section \ref{sec:stdreps}), and introduce new variables analogous to the effective resistances between nodes of the simplex (section \ref{sec:effres}).  To re-formulate the simplex integral in these variables, we  solve the inverse problem to express the original Schwinger parameters in terms of the effective resistances (section \ref{sec:reparam}). The re-parametrised integral, which will be the basis of our new shift operators,  then follows (section \ref{sec:CayMeng}).

\subsection{Graph polynomials} \label{sec:graphpolys}

Exponentiating all propagators via Schwinger parametrisation, the internal momenta can be integrated out reducing the simplex integral   to various scalar parametrisations.  The structure of the resulting Symanzik polynomials is clearest however when expressed in terms of the {\it inverse} of the usual 
variables.
For this reason, we use the inverse Schwinger parametrisation
\[\label{Schwingerintegral}
\frac{1}{q_{ij}^{2\alpha_{ij}+d}} = \frac{1}{\Gamma(\alpha_{ij}+d/2)}\int_0^\infty \D v_{ij} \,v_{ij}^{-d/2-\alpha_{ij}-1}e^{-q_{ij}^2/v_{ij}}.
\]
The resulting polynomials $\mathcal{U}$ and $\mathcal{F}$ are  then related to the standard Symanzik polynomials $U$ and $F$ by
\[\label{Kirchhofdef}
\mathcal{U}(v_{ij}) = \Big(\prod_{i<j}^n v_{ij}\Big) U\Big(\frac{1}{v_{ij}}\Big),
\qquad
\mathcal{F}(v_{ij}) = \Big(\prod_{i<j}^n v_{ij}\Big) F\Big( \frac{1}{v_{ij}}\Big).
\]
For the simplex, the  structure of  $\mathcal{U}$ and $\mathcal{F}$ can be expressed in terms of two matrices.  The first is the $(n-1)\times(n-1)$ Gram matrix $G_{ij}  = \bs{p}_i\cdot\bs{p}_j$.  For our purposes, the most convenient parametrisation is 
\[\label{Gramdef}
G_{ij} = \begin{cases} \sum_{k=1}^{n} V_{ik}\qquad &i=j \\
-V_{ij} \qquad & i\neq j
\end{cases}\
\qquad i,j = 1,\ldots, n-1,
\]
where 
\[\label{Vdef}
V_{ij} =\begin{cases} -\bs{p}_i\cdot \bs{p}_j \qquad & i\neq j\\
0\qquad & i=j
\end{cases}\qquad i,j = 1,\ldots, n.
\]
Here the $V_{ij}$  provide a full set of $n(n-1)/2$ symmetric and independent Lorentz invariants.
To write 
 the diagonal entries in the Gram matrix,  we used momentum conservation to express $p_i^2=-\sum_{k\neq i}^n \bs{p}_i\cdot\bs{p}_k$.
The second matrix  is simply the image of the Gram matrix under the mapping
$
V_{ij}\rightarrow v_{ij},
$ 
namely
\[\label{gdef}
g_{ij} = \begin{cases} \sum^{n}_{k=1} v_{ik} \qquad & i=j, \\
-v_{ij}\qquad & i\neq j, 
\end{cases}\qquad i,j=1,\ldots, n-1.
\]
Since the $v_{ij}$ correspond to the edges of the simplex we define,  as we did for the $V_{ij}$,
\[
v_{ij} = v_{ji}, \qquad v_{ii}=0.
\]

As shown in appendix \ref{incidence_app},
the graph polynomials are now  
\begin{empheq}[box=\nicebox]{align} 
\label{Kirchhoff1}
\mathcal{U}  = |g|, \qquad \mathcal{F}= \mathrm{tr}(\mathrm{adj}(g)\cdot G), \qquad \frac{\mathcal{F}}{\mathcal{U}} = \mathrm{tr}(g^{-1}\cdot G),
\end{empheq}
where $|g|=\mathrm{det}\,g$, $\mathrm{adj}\,g=|g| \,g^{-1}$ is the adjugate matrix and $g^{-1}$ the inverse matrix.  
The derivation proceeds by expressing the delta functions of \eqref{simplex} in Fourier form and integrating out all internal momenta.  Only  after this step has been performed are the Fourier integrals for the delta functions then  evaluated.
As the Gram determinant $|G|$ is proportional to the squared volume of the simplex spanned by the independent momenta, the polynomial $\mathcal{U}$ describes the image of this squared volume under the mapping $V_{ij}\rightarrow v_{ij}$.  Alternatively, by the matrix tree theorem (see {\it e.g.,} \cite{Weinzierl:2022eaz}), $\mathcal{U}$ is the Kirchhoff polynomial encoding the sum of  spanning trees on the simplex.

A second useful expression for $\mathcal{F}$ can be derived 
from  Jacobi's identity, 
\[\label{Jacobi}
\partial_{v_{ij}}|g| = \mathrm{tr}\big(\mathrm{adj}\,(g)\cdot \partial_{v_{ij}} g\big),
\] 
in combination with the relation
\[\label{bigGrel}
G_{ij} = \sum_{k<l}^n V_{kl} \frac{\partial G_{ij}}{\partial V_{kl}}= \sum_{k<l}^n V_{kl} \frac{\partial g_{ij}}{\partial v_{kl}}.
\]
This last relation  follows from the linearity of the $G_{ij}$ in the $V_{kl}$, as we saw in \eqref{Gramdef}, and the mapping of  $G_{ij}\rightarrow g_{ij}$ under $V_{kl}\rightarrow v_{kl}$.  The sums run over all $k$ and $l$ such that $k<l$, corresponding to all edges of the simplex. 
Substituting \eqref{bigGrel} into \eqref{Kirchhoff1} then using \eqref{Jacobi}, 
\[
\mathcal{F} = \sum_{i<j}^{n} \frac{\partial |g|}{\partial v_{ij}} \,V_{ij}  ,\qquad
\frac{\mathcal{F}}{\mathcal{U}} = \sum_{i<j}^{n}\frac{\partial\ln |g|}{\partial v_{ij}}\,V_{ij},
\]
or in terms of the raw momenta,
\begin{empheq}[box=\nicebox]{align} 
\label{Kirchhoff2}
\mathcal{F} = -\sum_{i<j}^{n}\frac{\partial |g|}{\partial v_{ij}}\,  \bs{p}_i\cdot\bs{p}_j,\qquad
\frac{\mathcal{F}}{\mathcal{U}} = -\sum_{i<j}^{n}\frac{\partial\ln |g|}{\partial v_{ij}}\,  \bs{p}_i\cdot\bs{p}_j.
\end{empheq}

\subsection[\texorpdfstring{Parametric representations of the
$n$-point correlator}{Parametric representations of the n-point correlator}]{\texorpdfstring{Parametric representations of the
\boldmath{$n$}-point correlator}{Parametric representations of the n-point correlator}}\label{sec:stdreps}

To express correlators compactly,  we extract the overall delta function of momentum conservation as
\[\label{redcorr}
\< \O_1(\bs{p}_1) \ldots \O_n(\bs{p}_n) \> = \lla \O_1(\bs{p}_1) \ldots \O_n(\bs{p}_n) \rra (2\pi)^d\delta(\sum_{i=1}^n \bs{p}_n).
\]
We also define an arbitrary function $f(\hat{\bs{v}})$ whose arguments, denoted collectively by the vector $\hat{\bs{v}}$, are the independent  inverse Schwinger parameter cross ratios
\[\label{vcrossratio}
v_{[ijkl]} =  \frac{v_{ij} v_{kl}}{v_{ik}v_{jl}}.
\]  
The simplex integral \eqref{simplex} can now be written in 
a variety of standard forms using the polynomials $\mathcal{U}$ and $\mathcal{F}$ defined in \eqref{Kirchhoff1} or \eqref{Kirchhoff2}.  Among the most useful  are: 
\begin{itemize}
\item[1.] {\it Schwinger parametrisation:}
\[\label{Schwrep}
\lla \O_1(\bs{p}_1) \ldots \O_n(\bs{p}_n) \rra = \Big(\prod_{i<j}^n\int_0^\infty \D v_{ij}\, v_{ij}^{-\alpha_{ij}-1}\Big) f(\hat{\bs{v}})\,\mathcal{U}^{-d/2} e^{-\mathcal{F}/\mathcal{U}}
\]
Here, the $v_{ij}^{-d/2}$ factors in \eqref{Schwingerintegral} cancel with those associated with $\mathcal{U}^{-d/2}$ via \eqref{Kirchhofdef}.
\item[2.] {\it Lee-Pomeransky  parametrisation} \cite{Lee:2013hzt}:
\[\label{LPrep}
\lla \O_1(\bs{p}_1) \ldots \O_n(\bs{p}_n) \rra =\Big(\prod_{i<j}^n\int_0^\infty \D v_{ij}\, v_{ij}^{-\alpha_{ij}-1} \Big)f(\hat{\bs{v}})\,(\mathcal{U}+\mathcal{F})^{-d/2}
\]
\item[3.] {\it Feynman  parametrisation:}
\[\label{Feynrep}
\lla \O_1(\bs{p}_1) \ldots \O_n(\bs{p}_n) \rra =\Big(\prod_{i<j}^n\int_0^\infty \D v_{ij}\,v_{ij}^{-\alpha_{ij}-1} \Big)\delta\Big(1-\sum_{i<j}^n \kappa_{ij}v_{ij}\Big) f(\hat{\bs{v}})\,\mathcal{U}^{\omega-d/2} \mathcal{F}^{-\omega}
\]
where $\omega=(n-1)d/2+\sum_{i<j}^n\alpha_{ij} = (-\Delta_t+(n-1)d)/2$ and the constants $\kappa_{ij}\ge 0$ can be chosen arbitrarily provided they are not all zero.\footnote{
The Feynman parametrisation follows from the Schwinger parametrisation by setting $v_{ij}=y_{ij}/\sigma$ subject to the constraint $\sum_{i<j}^n \kappa_{ij}y_{ij}=1$. The  $\mathcal{U}$ and $\mathcal{F}$ are homogeneous polynomials of weights $(n-1)$ and $(n-2)$ respectively, meaning that $\mathcal{F}(v_{ij})/\mathcal{U}(v_{ij}) =\sigma\mathcal{F}(y_{ij})/\mathcal{U}(y_{ij})$ while the Jacobian can be evaluated as per appendix B of \cite{Bzowski:2020kfw}.  We then perform the scale integral over $\sigma$ and relabel the $y_{ij}\rightarrow v_{ij}$. }
If we choose all $\kappa_{ij}=1$ then the integration region is a simplex in the space spanned by the $v_{ij}$.  Alternatively, we can set a single $\kappa_{ij}$ to unity and the rest to zero which trivialises one of the integrations at the cost of obscuring permutation invariance.

\end{itemize}

These representations are all equivalent up to numerical factors; for clarity, we  have re-absorbed these into the arbitrary functions.  
For analysing the action of weight-shifting operators and verifying the conformal Ward identities, we will focus exclusively on the Schwinger parametrisation \eqref{Schwrep}.  Nevertheless, the Lee-Pomeransky representation \eqref{LPrep} is well suited for studying the Landau singularities, as discussed in appendix \ref{Landau_app}, and the Feynman parametrisation \eqref{Feynrep} has the  virtue that 
 one integral can be performed using the delta function.

\paragraph{Example:} As a quick illustration, the $4$-point function in Schwinger parametrisation is
\begin{align}
&\lla \O_1(\bs{p}_1)  \O_2(\bs{p}_2) \O_3(\bs{p}_3)\O_4(\bs{p}_4) \rra \nn\\& \qquad = \Big(\prod_{i<j}^4\int_0^\infty \D v_{ij}\, v_{ij}^{-\alpha_{ij}-1}\Big) f\Big(\frac{v_{12}v_{34}}{v_{13}v_{24}},\,\frac{v_{14}v_{23}}{v_{13}v_{24}}\Big)\,|g|^{-d/2} e^{-\mathrm{tr}\,( g^{-1}\cdot G)}
\end{align}
where $G_{ij} = \bs{p}_i\cdot \bs{p}_j$ is the  $3\times 3$ Gram matrix 
and $g$ is its image
\begin{align}
g =\left( \begin{matrix}v_{12}+v_{13}+v_{14} & -v_{12} & -v_{13}\\ -v_{12} & v_{12}+v_{23}+v_{24} & -v_{23}\\
-v_{13} & -v_{23} & v_{13}+v_{23}+v_{34}\end{matrix}\right).
\end{align} 
The determinant is
\begin{align}\label{4ptU}
|g| &=  v_{12} v_{13} v_{14} + v_{12} v_{14} v_{23} + v_{13} v_{14} v_{23} + v_{12} v_{13} v_{24} \nn\\&\quad + v_{13} v_{14} v_{24} + 
 v_{12} v_{23} v_{24} + v_{13} v_{23} v_{24} + v_{14} v_{23} v_{24}\nn\\&\quad + v_{12} v_{13} v_{34} + v_{12} v_{14} v_{34} +
  v_{12} v_{23} v_{34} + v_{13} v_{23} v_{34} \nn\\&\quad + v_{14} v_{23} v_{34} + v_{12} v_{24} v_{34} + 
 v_{13} v_{24} v_{34} + v_{14} v_{24} v_{34}
\end{align}
and the equivalence of \eqref{Kirchhoff1} and \eqref{Kirchhoff2} can be verified directly.

\subsection{The effective resistances}\label{sec:effres}

Thus far, we have expressed the Kirchhoff polynomial
$\mathcal{U}$ as the determinant of $g$, the image under $V_{ij}\rightarrow v_{ij}$ of the 
 Gram matrix, where $\bs{p}_n$ is eliminated using momentum conservation.
However, since all vertices of the simplex are equivalent, $\mathcal{U}$ ought also to be expressible in terms of the $n\times n$ matrix $\tilde{g}$ corresponding to the image 
of the {\it extended} Gram matrix $\tilde{G}_{ij} = \bs{p}_i\cdot\bs{p}_j$ for $i,j=1,\ldots,n$.  
This is simply the Laplacian matrix for the simplex: 
\[\label{tildeg}
\tilde{g}_{ij} = \begin{cases} \sum^{n}_{k=1} v_{ik}, \qquad & i=j, \\
-v_{ij},\qquad & i\neq j, 
\end{cases}\qquad i,j=1,\ldots, n.
\]
As every row and column sum of the Laplacian matrix is zero its determinant 
 vanishes identically,  
but its cofactors ({\it i.e.,} signed first minors) 
are in fact all equal to $\mathcal{U}$.
To see this, consider 
the diagonal minor 
 $|\tilde{g}^{(n,n)}|$  formed by deleting row $n$ and column $n$ then taking the determinant.   Comparing with \eqref{gdef}, we then see that  $|\tilde{g}^{(n,n)}|=|g|=\mathcal{U}$.  As any diagonal minor is equal to its cofactor, $\mathcal{U}$ is likewise the $(n,n)$ cofactor.   However, by elementary row and column operations one can show that all cofactors of the Laplacian matrix are equal.\footnote{For example, add one to every element of $\tilde{g}_{ij}$ then add all rows to the first row, and all columns to the first column.  The top left entry is now $n^2$ while all remaining entries of the first row and column are $n$. Taking the determinant, we first extract an overall factor of $n$ from the top row, then subtract the new top row (whose leftmost entry is now $n$ with all other entries one) from all the other rows. The resulting matrix has zeros in all entries of the first column apart from the top one which is $n$, and all entries other than those in the first row and column are $\tilde{g}_{ij}$ (since we added one then subtracted one).  The determinant of $\tilde{g}_{ij}$ plus the all ones matrix is therefore $n^2$ times the $(1,1)$ cofactor of $\tilde{g}_{ij}$.  Repeating the exercise for any other choice of row and column yields the same result with the corresponding cofactor, hence all cofactors are equal.  Note this also shows that  $\mathcal{U}$ is $n^{-2}$ times the determinant of the Laplacian plus the all-ones matrix.}  
Thus, every cofactor (and every diagonal minor) is equal to  $\mathcal{U}$.  Note this also confirms that our choice of eliminating $\bs{p}_n$ in section \ref{sec:graphpolys} was immaterial.

Let us now turn to an electrical analogy involving a simplicial network of resistors.   Here, the Laplacian matrix naturally encodes the external current $\mathcal{I}_i$ 
flowing into node $i$, since 
\[
\mathcal{I}_i = \sum_{j\neq i} v_{ij}(\mathcal{V}_i-\mathcal{V}_j) = \sum_{j\neq i} \tilde{g}_{ij}\mathcal{V}_j,
\] 
where  $v_{ij}$ is the conductivity of the edge connecting nodes $i$ and $j$ and $\mathcal{V}_j$ is the voltage of node $j$. 
Given this identification of the $v_{ij}$ with the conductivities, a natural question to ask is what are the corresponding {\it effective resistances} 
between the nodes?   From Kirchhoff, the effective resistance $s_{ij}$ between nodes $i$ and $j$ is given by the ratio of minors \cite{kirchhoff1847,kirchhoff1958}
\[
s_{ij} = \frac{|\tilde{g}^{(ij,ij)}|}{|\tilde{g}^{(j,j)}|}, 
\]
where $|\tilde{g}^{(I,J)}|$ indicates the minor formed by deleting the set of rows $I$ and columns $J$ then taking the determinant.
Thus,  $|\tilde{g}^{(ij,ij)}|$ is the second minor formed by  deleting rows $i$ and $j$ as well as columns $i$ and $j$, while $|\tilde{g}^{(j,j)}|$ is  the first minor corresponding to deleting row and column $j$.
From \eqref{tildeg}, the element $v_{ij}$ appears only in the row and columns $(i,i)$, $(i,j)$, $(j,i)$ and $(j,j)$  of $\tilde{g}$.  Forming the  first minor $|\tilde{g}^{(j,j)}|$ by deleting row and column $j$, $v_{ij}$ then appears only once in the $(i,i)$ position. 
The derivative $\partial |\tilde{g}^{(j,j)}|/\partial v_{ij}$ is thus equal to the second minor $|\tilde{g}^{(ij,ij)}|$ formed by additionally deleting row and column $i$ in $|\tilde{g}^{(j,j)}|$.
Since $|\tilde{g}^{(j,j)}|=|g|$ as above, we have
\begin{empheq}[box=\nicebox]{align} 
\label{sdef}
s_{ij} = \frac{\partial \ln |g|}{\partial v_{ij}}, \qquad \frac{\mathcal{F}}{\mathcal{U} }= -\sum_{i<j}^n s_{ij}\,\bs{p}_i\cdot\bs{p}_j,
\end{empheq}
where the second result  follows immediately from \eqref{Kirchhoff2}.
The Schwinger exponent in \eqref{Schwrep} thus  encodes the effective resistances $s_{ij}$ between all vertices.  Moreover, both  $\mathcal{U}$ and $\mathcal{F}$ have been related to minors of the Laplacian: $\mathcal{U}$ is any diagonal first minor (or cofactor), while the coefficients of the $\mathcal{F}$ polynomial correspond to the second minors: from \eqref{Kirchhoff2}, the coefficient of $V_{ij}=-\bs{p}_i\cdot\bs{p}_j$ (for $i<j$) is  $
\partial |g|/\partial v_{ij} = |\tilde{g}^{(ij,ij)}|
$. 

Earlier, we noted that $\mathcal{U}=|g|$ is proportional to the squared volume of the $(n-1)$-simplex formed by the 
 independent momenta under the map $V_{ij}\rightarrow v_{ij}$.
By the same token, each coefficient $|\tilde{g}^{(ij,ij)}|$ of the $\mathcal{F}$ polynomial thus corresponds to the image of  $|\tilde{G}^{(ij,ij)}|$, the second minor of the extended Gram matrix.
However, this minor is simply the determinant of the reduced Gram matrix formed from all the momenta apart from $\bs{p}_i$ and $\bs{p}_j$.
Thus, the coefficient of $V_{ij}$ in the $\mathcal{F}$ polynomial
is 
proportional to the squared volume of the $(n-2)$-simplex, formed from all the momenta except for $\bs{p}_i$ and $\bs{p}_j$, under the map $V_{ij}\rightarrow v_{ij}$. 
Similarly, the effective resistance $s_{ij}$ is proportional to the ratio of the squared volume of this $(n-2)$-simplex to the squared volume of the full $(n-1)$-simplex.

\subsection{Re-parametrising the simplex}
\label{sec:reparam}

The original Schwinger parametrisation \eqref{Schwrep} is complicated by the non-linear dependence of  the exponent on the $v_{ij}$.   As we  saw in \eqref{sdef},  however, the coefficients of the $V_{ij}=-\bs{p}_i\cdot\bs{p}_j$ 
in $\mathcal{F}/\mathcal{U}$ are simply the effective resistances $s_{ij}$ between nodes.
The next step  is thus to invert the relation \eqref{sdef} to find the $v_{ij}$ in terms of the $s_{ij}$, {\it i.e.,} to express the conductivities in terms of the effective resistances. 
The simplex integral can then be fully re-parametrised in terms of the $s_{ij}$, with the linearity of the Schwinger exponent giving a Fourier-style duality between the $V_{ij}$ and the $s_{ij}$.  This duality means that all momentum derivatives acting on the simplex, and all momenta, can be trivially exchanged for operators constructed from the $s_{ij}$ and derivatives $\partial/\partial s_{ij}$.  The latter can then be integrated by parts.  This strategy will repeatedly prove useful to us later.

We start by applying Jacobi's relation to further evaluate \eqref{sdef},
\begin{align}
s_{ij}  =\frac{1}{|g|}\frac{\partial|g|}{\partial v_{ij}}=\mathrm{tr}\Big(g^{-1}\cdot\frac{\partial g}{\partial v_{ij}}\Big)=\begin{cases}(g^{-1})_{ii}+(g^{-1})_{jj}-2(g^{-1})_{ij}, \qquad &i<j< n \\ (g^{-1})_{ii},\qquad &i< j=n\end{cases}\label{srel1}
\end{align}
where the matrices $\partial g/\partial v_{ij}$ are easily evaluated from \eqref{gdef}. 
Defining $s_{ii}=0$  for convenience (as we similarly defined $v_{ii}=0$) and re-arranging, we find 
\[\label{ginvs}
(g^{-1})_{ij} = \frac{1}{2}(s_{in}+s_{jn}-s_{ij}), \qquad i,j=1,\ldots, n-1
\]
where the diagonal entries reduce to $(g^{-1})_{ii}=s_{in}$.
Inverting this matrix will now give us back the matrix $g$, as defined in \eqref{gdef}, but re-expressed in terms of the $s_{ij}$.  The  desired expressions for the $v_{ij}$ in terms of the $s_{ij}$ can then be read off from the appropriate entries.

In fact,  it is sufficient simply to know the determinant $|g^{-1}|$.  For $i<j<n$, the $(i,j)$ minor formed by deleting row $i$ and column $j$ of $g^{-1}$ 
is $|(g^{-1})^{(i,j)}|=-(-1)^{i+j}\partial |g^{-1}|/\partial s_{ij}$, since from \eqref{ginvs} $s_{ij}$ appears (with coefficient minus one-half) only in the positions $(i,j)$ and $(j,i)$ of the symmetric matrix $g^{-1}$.  
The off-diagonal entries of the adjugate matrix are thus
\[
\mathrm{adj}(g^{-1})_{ij} = (-1)^{i+j}|(g^{-1})^{(i,j)}| = - \frac{\partial |g^{-1}|}{\partial s_{ij}}, \qquad i<j<n
\] 
so 
\[\label{vfromginv}
v_{ij} = -g_{ij} = -\frac{1}{|g^{-1}|}\,\mathrm{adj}(g^{-1})_{ij} = \frac{\partial \ln |g^{-1}|}{\partial s_{ij}}, \qquad i<j<n.
\]
Similarly,  $s_{in}$ appears in every entry of the $i^{\mathrm{th}}$ row of $g^{-1}$, and in every entry of the $i^{\mathrm{th}}$ column.
The coefficients for the off-diagonal entries are all one-half, while that for the diagonal entry is one.
The derivative $\partial |g^{-1}|/\partial s_{in}$ then corresponds to summing one-half times the signed minors  both along the $i^{\mathrm{th}}$ row and down the $i^{\mathrm{th}}$ column such that the diagonal entry is counted twice.  As $g^{-1}$ is symmetric, however, these two sums are equal so we can simply sum along the $i^{\mathrm{th}}$ row only with coefficient one.
This gives
\[\label{vfromginvn}
\frac{\partial \ln |g^{-1}|}{\partial s_{in}} = \sum_{j=1}^{n-1}\frac{ (-1)^{i+j}}{|g^{-1}|}|(g^{-1})^{(i,j)}|=\sum_{j=1}^{n-1}\frac{1}{|g^{-1}|}\mathrm{adj} (g^{-1})_{ij} = \sum_{j=1}^{n-1} g_{ij} = v_{in},
\]
where in the final step we used \eqref{gdef} to identify the sum of the first $n-1$ entries along the $i^{\mathrm{th}}$ row of the Laplacian as $v_{in}$.   The relation \eqref{vfromginv} thus holds for all $i<j\le n$.

To simplify this formula further, we observe that  $|g^{-1}|$ can be re-expressed in terms of the determinant of the $(n+1)\times (n+1)$ Cayley-Menger matrix,
\[\label{mdef}
m = \left(\begin{matrix}
 0 & s_{12} & s_{13}  & \ldots & s_{1n} & 1\\
 s_{12} & 0 & s_{23}  & \ldots & s_{2n} & 1\\
 s_{13} & s_{23} & 0  & \ldots & s_{3n} & 1\\ 
  \vdots & \vdots & \vdots & &\vdots&\vdots\\
  s_{1n} & s_{2n} & s_{3n}  & \ldots & 0 & 1\\
 1 & 1 & 1&\ldots & 1& 0\\
\end{matrix}\right).
\]
When evaluating the determinant, if we subtract the $n^{\mathrm{th}}$ column from the first $(n-1)$  columns, and then the $n^{\mathrm{th}}$ row from the first $(n-1)$  rows, we find  
\[
|m| = \left|\begin{matrix}
 -2s_{1n}\quad  & s_{12}-s_{1n}-s_{2n}\quad  & s_{13}-s_{1n}-s_{3n} & \ldots & s_{1n} \quad & 0 & \\[1ex]
\,\,  s_{12} -s_{1n}-s_{2n}& -2s_{2n} & s_{23}-s_{2n}-s_{3n}  & \ldots & s_{2n} & 0\\[1ex]
 s_{13} -s_{1n}-s_{3n} & s_{23}-s_{2n}-s_{3n} & -2s_{3n}  & \ldots & s_{3n} & 0\\[1ex] 
  \vdots & \vdots & \vdots & &\vdots&\vdots\\[1ex]
  s_{1n} & s_{2n} & s_{3n}  & \ldots & 0 & 1\\[1ex]
 0 & 0 & 0&\ldots & 1& 0\\
\end{matrix}\right|.
\]
Comparing with \eqref{ginvs}, the upper-left $(n-1)\times (n-1)$ submatrix is $-2g^{-1}$.
Laplace expanding along the $(n+1)^\mathrm{th}$ row and then the $(n+1)^\mathrm{th}$ column thus gives
\[\label{mginvreln}
|m| = -(-2)^{n-1}|g^{-1}|.
\]
Equations \eqref{vfromginv} and \eqref{vfromginvn} can now be  cleanly re-expressed in terms of the Cayley-Menger determinant:
\begin{empheq}[box=\nicebox]{align} 
\label{vdef}
v_{ij} = \frac{\partial \ln |m|}{\partial s_{ij}}, \qquad i<j\le n.
\end{empheq}
This is our desired result expressing all the $v_{ij}$ in terms of the $s_{ij}$, inverting \eqref{sdef}.
A few additional relations also follow.  
 Jacobi's relation allows us to write 
\[
v_{ij} =\frac{1}{|m|} \frac{\partial  |m|}{\partial s_{ij}} =\frac{1}{|m|}\mathrm{tr}\Big(\mathrm{adj}(m)\cdot\frac{\partial m}{\partial s_{ij}}\Big)=\mathrm{tr}\Big(m^{-1}\cdot\frac{\partial m}{\partial s_{ij}}\Big) =  2(m^{-1})_{ij}, \label{vrel1}
\]
since  $\partial m_{kl}/\partial s_{ij} = 2\delta_{i(k}\delta_{l)j}$ from \eqref{mdef}.
As the off-diagonal entries of the Laplacian matrix are $\tilde{g}_{ij}=-v_{ij}$, this means that
\[\label{minvisLa1}
\tilde{g}_{ij} = -2 (m^{-1})_{ij}, \qquad  i,j\le n.
\]
In fact, as indicated, this equation also holds for the diagonal elements with $i=j\le n$, since if we multiply the $(n+1)^{\mathrm{th}}$ row of $m$ by column $i$ of $m^{-1}$ we find
\[
0 = \sum_{j=1}^n m^{-1}_{ij}, \qquad i\le n
\]
and since all row and column sums of the Laplacian matrix vanish, 
\[\label{minvisLa2}
\tilde{g}_{ii} = -\sum_{j\neq i}^n \tilde{g}_{ij} = \sum_{j\neq i}^n 2(m^{-1})_{ij} = -2(m^{-1})_{ii}.
\]
Thus, the $n\times n$ upper-left  submatrix of the inverse Cayley-Menger matrix is minus one-half the Laplacian matrix,
using either \eqref{sdef} or \eqref{vdef} to convert between the $s_{ij}$ and $v_{ij}$.\footnote{
To the best of our knowledge, this result, 
along with a geometrical interpretation of the  remaining $(n+1)^{\mathrm{th}}$ row and column of the Cayley-Menger inverse, was first obtained by Fiedler, see \cite{fiedler_2011, Devriendt_2022}.}

The appearance of the Cayley-Menger matrix in our analysis is not a total surprise: in Euclidean distance geometry, the Cayley-Menger determinant is proportional to the squared volume of the simplex whose squared side lengths are given by the $s_{ij}$.  Here, the map $v_{ij}\rightarrow V_{ij}$ sends $(g^{-1})_{ij}$ to the inverse Gram matrix $G^{-1}_{ij} = (\tilde{\bs{p}}_i\cdot\tilde{\bs{p}}_j)$, which is itself the Gram matrix formed from the independent {\it dual} momentum vectors $\tilde{\bs{p}}_i=G^{-1}_{ij}\bs{p}_j$ satisfying $\tilde{\bs{p}}_i\cdot \bs{p}_j=\delta_{ij}$.  
The determinant $|g^{-1}|$ is thus proportional to the squared volume of the dual $(n-1)$-simplex spanned by the independent $\tilde{\bs{p}}_i$, and
by \eqref{mginvreln}, the $s_{ij}$ are then the squared side lengths of this dual simplex. 
This provides an alternative (dual) geometrical interpretation for the $s_{ij}$, besides the volume ratio discussed at the end of section \ref{sec:effres}.

\paragraph{Example:}  All the relations above are easily checked for  small values of $n$, and the $s_{ij}$ are always rational functions of the $v_{ij}$ and vice versa.
For the 4-point function, we find, {\it e.g.,} 
\begin{align}
s_{12}&= \frac{\partial \ln |g|}{\partial v_{12}} = ( v_{13} v_{14} + v_{14} v_{23} + v_{13} v_{24} + v_{23} v_{24} + v_{13} v_{34} + v_{14} v_{34} + v_{23} v_{34} + 
 v_{24} v_{34})\, |g|^{-1} \nn\\[1ex]
v_{12}&= \frac{\partial \ln |m|}{\partial s_{12}} =(-2 s_{13} s_{23} + 2 s_{14} s_{23} + 2 s_{13} s_{24} - 2 s_{14} s_{24} - 4 s_{12} s_{34} 
\nn\\[-0.5ex]&\qquad\qquad\qquad 
+  2 s_{13} s_{34} 
 + 2 s_{14} s_{34} + 2 s_{23} s_{34} + 2 s_{24} s_{34} - 2 s_{34}^2)\,|m|^{-1},
\end{align}
where  $|g|$ was evaluated in \eqref{4ptU} and 
\begin{align}\label{m4}
|m| &= \left|\,\begin{matrix}
  0 & s_{12} & s_{13}  & s_{14} & 1\\
 s_{12} & 0 & s_{23}  & s_{24} & 1\\
 s_{13} & s_{23} & 0  & s_{34} & 1\\ 
  s_{14} & s_{24} & s_{34}  & 0 & 1\\
 1 & 1 & 1 & 1& 0
\end{matrix}\,\right|
\end{align}
evaluates to 
\begin{align}\label{detm4}
|m| &=-2 s_{12} s_{13} s_{23} + 2 s_{12} s_{14} s_{23} + 2 s_{13} s_{14} s_{23} - 2 s_{14}^2 s_{23}
 -  2 s_{14} s_{23}^2 + 2 s_{12} s_{13} s_{24} - 2 s_{13}^2 s_{24} 
 \nn\\
&\quad 
 - 2 s_{12} s_{14} s_{24} + 
 2 s_{13} s_{14} s_{24} + 2 s_{13} s_{23} s_{24} + 2 s_{14} s_{23} s_{24} - 2 s_{13} s_{24}^2 - 
 2 s_{12}^2 s_{34} + 2 s_{12} s_{13} s_{34}
\nn\\
&\quad 
  + 2 s_{12} s_{14} s_{34} - 2 s_{13} s_{14} s_{34} + 
 2 s_{12} s_{23} s_{34} + 2 s_{14} s_{23} s_{34} + 2 s_{12} s_{24} s_{34} + 2 s_{13} s_{24} s_{34} 
\nn\\
&\quad 
 -  2 s_{23} s_{24} s_{34} - 2 s_{12} s_{34}^2.
 \end{align}

\subsection{\texorpdfstring{Cayley-Menger parametrisation of the $n$-point correlator}{Cayley-Menger parametrisation of the n-point correlator}}\label{sec:CayMeng}

Using the results above, we can re-express the various 
parametrisations of the 
simplex integral in terms of the effective resistances $s_{ij}$.  
If we write the external momenta in Cayley-Menger form,
\[
M = m\Big|_{s_{ij}\rightarrow \bs{p}_i\cdot \bs{p}_j}
\]
the Schwinger exponent can be written as
\[\label{CaySchwexp}
-\frac{\mathcal{F}}{\mathcal{U}} = \sum_{i<j}^n s_{ij}(\bs{p}_i\cdot\bs{p}_j) = \frac{1}{2}\mathrm{tr} (M\cdot m) + n,
\]
where the constant term $n$ just produces an overall scaling which can be re-absorbed into the arbitrary function of cross-ratios. 
Moreover, as shown in appendix \ref{Jac_app}, the 
determinant of the Jacobian is 
\[\label{Jac0}
\left|\frac{\partial s}{\partial v} \right| = \left|\frac{\partial^2 \ln |g|}{\partial v\partial v}\right|\propto |g|^{-n} \propto |m|^{n},
\]
where the constant of proportionality can again be absorbed into the arbitrary function.

The Schwinger form \eqref{Schwrep} of  the simplex integral  now  becomes
\[\label{Cay1}
\lla \O_1(\bs{p}_1) \ldots \O_n(\bs{p}_n) \rra  = \Big(\prod^n_{i<j} \int_0^\infty\mathrm{d} s_{ij} \, \Big(\frac{\partial \ln |m|}{\partial s_{ij}}\Big)^{-\alpha_{ij}-1} \Big)f(\hat{\bs{v}}) \,|m|^{d/2-n}\, e^{\frac{1}{2}\mathrm{tr} (M\cdot m)}
\]
where the cross-ratios $\hat{\bs{v}}$ are rational functions of the $s_{ij}$ as defined via \eqref{vcrossratio} and \eqref{vdef}.
An alternative expression can be given in terms of the Cayley-Menger minors, since
from Jacobi's relation
\[\label{dmdstominnors}
\frac{\partial |m|}{\partial s_{ij}} = 2(-1)^{i+j}\,|m^{(i,j)}|,
\]
where $|m^{(i,j)}|$ is the minor formed by taking the determinant after deleting row $i$ and column $j$.
After absorbing numerical factors into the arbitrary function, this gives
\[\label{Cay2}
\lla \O_1(\bs{p}_1) \ldots \O_n(\bs{p}_n) \rra  = \Big(\prod^n_{i<j} \int_0^\infty\mathrm{d} s_{ij} \, |m^{(i,j)}|^{-\alpha_{ij}-1} \Big)\,f(\hat{\bs{v}})\,|m|^{\alpha}\, e^{\frac{1}{2}\mathrm{tr} (M\cdot m)}
\]
where 
\[\label{alphadef}
\alpha=\frac{1}{2}\Big(d+n(n-3)-\sum_{i=1}^n\Delta_i\Big)
,\qquad
\hat{\bs{v}}_{[ijkl]} = 
\frac{v_{ij}v_{kl}}{v_{ik}v_{jl}} = \frac{|m^{(i,j)}| |m^{(k,l)}|}{|m^{(i,k)}||m^{(j,l)}|}.
\]
Analogous expressions can be obtained for the Lee-Pomeransky and Feynman representations \eqref{LPrep} and \eqref{Feynrep}, but the Schwinger parametrisations \eqref{Cay1} and \eqref{Cay2} are particularly convenient. 
As noted, 
the diagonal Schwinger exponent 
means differential operators in the momenta can easily be traded for equivalent differential operators in the $s_{ij}$ acting on the exponential, whose action can be further evaluated through integration by parts.

\section{Weight-shifting operators}
\label{sec:wsops}

New weight-shifting operators now follow 
from the Cayley-Menger parametrisation \eqref{Cay2}.  Acting on the Schwinger exponent \eqref{CaySchwexp} with an appropriate polynomial differential operator in the momenta pulls down a corresponding polynomial in the $s_{ij}$.
Choosing these polynomials to be the Cayley-Menger determinant and its minors, we obtain shift operators either increasing $\alpha$ or decreasing one of the $\alpha_{ij}$ by integer units. 
We discuss these new operators in section \ref{sec:dup2ops}, showing their effect is to increase the spacetime dimension by two while performing assorted shifts of the operator dimensions. 
 Further weight-shifting operators can then be constructed by conjugating these operators with shadow transforms as shown in section \ref{shadowsec}.
Explicit examples  are given for the $3$- and $4$-point functions in section \ref{3n4ptShifts}. 
 Then, in section \ref{sec:Wops}, we  turn to analyse the weight-shifting operators proposed in \cite{Karateev:2017jgd}.   These preserve the spacetime dimension but  their action can nevertheless be understood using our parametric representations.

\subsection[\texorpdfstring{New operators sending $d\rightarrow d+2$}{New operators sending d to d+2}]{\texorpdfstring{New operators sending \boldmath{$ d\rightarrow d+2$}}{New operators sending d to d+2}}
\label{sec:dup2ops}

Let us begin with the $V_{ij}$ defined in \eqref{Vdef} as our independent momentum variables. 
Acting on the Schwinger exponent \eqref{CaySchwexp}, for any $i<j$ 
\[
-\frac{\partial}{\partial V_{ij} }e^{\frac{1}{2}\mathrm{tr}(M\cdot m)} = s_{ij}\,e^{\frac{1}{2}\mathrm{tr}(M\cdot m)}, \qquad  
 -V_{ij} \,e^{\frac{1}{2}\mathrm{tr}(M\cdot m)} = \frac{\partial}{\partial s_{ij}}e^{\frac{1}{2}\mathrm{tr}(M\cdot m)}
\]
allowing differential operators in the momenta to be traded for  equivalent operators in the integration variables 
$s_{ij}$.
The shift operators 
\begin{empheq}[box=\nicebox]{align} 
\label{Sdef}
\mathcal{S}_{ij}^{++} = |m^{(i,j)}|\Big|_{s_{ij}\rightarrow -\partial/\partial V_{ij}}, \qquad \mathcal{S} = |m|\Big|_{s_{ij}\rightarrow -\partial/\partial V_{ij}},
\end{empheq}
then serve to pull down factors of $|m^{(i,j)}|$ and $|m|$ respectively, thus their action is to send 
\[
\mathcal{S}_{ij}^{++}: \quad \alpha_{ij}\rightarrow \alpha_{ij} -1, \qquad
\mathcal{S}:\quad \alpha\rightarrow \alpha+1.
\]
From \eqref{alphaijdef} and \eqref{alphadef}, this is equivalent to shifting
\begin{align}\label{Sijshifts}
\mathcal{S}_{ij}^{++}: \quad &d\rightarrow d+2,\quad \Delta_i\rightarrow \Delta_i+1,\quad
\Delta_j\rightarrow \Delta_j+1,\\\mathcal{S}:\quad & d\rightarrow d+2,
\end{align}
and so the superscript on $\mathcal{S}_{ij}^{++}$ is chosen to indicate its action of raising $\Delta_i$ and $\Delta_j$ by one. 

While the Cayley-Menger structure of $\mathcal{S}_{ij}^{++}$ and $\mathcal{S}$ is manifest in the variables $V_{ij}$, where convenient these operators can easily be rewritten in terms of other scalar invariants ({\it e.g.,} Mandelstam variables) via the chain rule.  We will discuss this for 3- and 4-point functions shortly in section \ref{3n4ptShifts}.  

Alternatively, we can  express  $\mathcal{S}_{ij}^{++}$ and $\mathcal{S}$   in terms of {\it vectorial} derivatives with respect to independent momentum  $\bs{p}_i$ for $i=1,\ldots n-1$.  For $\mathcal{S}$, we find 
\begin{empheq}[box=\nicebox]{align} \label{Sdef2}
\mathcal{S} = -\, \frac{(n-1)!}{|G|}\, p_{1}^{[\mu_1}\ldots p_{n-1}^{\mu_{n-1]}}\frac{\partial}{\partial p_1^{\mu_1}}\ldots \frac{\partial}{\partial p_{n-1}^{\mu_{n-1}}}
\end{empheq}
where 
$|G|=|\bs{p}_i\cdot\bs{p}_j|$ is the Gram determinant and the $\mu_i$ are Lorentz indices.  (We leave all Lorentz indices upstairs to avoid confusion with the momentum labels, given we are working on a flat Euclidean metric.) 
The equivalence of \eqref{Sdef2} to \eqref{Sdef} can established either by direct calculation for specific $n$, or else by considering its action on the Schwinger exponential of the representation \eqref{Schwrep}.  
This representation is the appropriate one since, from \eqref{Kirchhoff1}, it involves only dot products of the {\it independent} momenta. 
Evaluating,  we find
\begin{align}\label{Scalc}
&\mathcal{S}\,\Big( e^{-\sum_{i,j}^{n-1} (g^{-1})_{ij}\bs{p}_i\cdot \bs{p}_j }\Big) 
 \nn\\&\quad
=-\frac{(n-1)!}{|G|}\, p_1^{[\mu_1}\ldots p_{n-1}^{\mu_{n-1}]}
\nn\\&\qquad\times
\Big(-2\sum_{j_1}^{n-1}(g^{-1})_{1{j_1}}p_{j_1}^{\mu_1}\Big)\ldots
\Big(-2\sum_{j_{n-1}}^{n-1}(g^{-1})_{n-1,{j_{n-1}}}p_{j_{n-1}}^{\mu_{n-1}}\Big)e^{-\sum_{i,j}^{n-1} (g^{-1})_{ij}\bs{p}_i\cdot \bs{p}_j }\nn\\
%
%
%
%
&\quad =
\frac{-(-2)^{n-1}(n-1)!}{|G|}\,\sum_{j_1,k_1}^{n-1}\ldots \sum_{j_{n-1},k_{n-1}}^{n-1}(g^{-1})_{1j_1}\ldots (g^{-1})_{n-1,j_{n-1}}(\bs{p}_1\cdot\bs{p}_{k_1})\ldots (\bs{p}_{n-1}\cdot\bs{p}_{k_{n-1}})
\nn\\[0ex]&\quad \qquad \qquad\qquad \qquad  \qquad\qquad \qquad \times\delta^{[k_1}_{j_1}\ldots \delta^{k_{n-1}]}_{j_{n-1}}e^{-\sum_{i,j}^{n-1} (g^{-1})_{ij}\bs{p}_i\cdot \bs{p}_j }\nn\\[1ex]
&\quad=-(-2)^{n-1}|g|^{-1}e^{-\sum_{i,j}^{n-1} (g^{-1})_{ij}\bs{p}_i\cdot \bs{p}_j }
\nn\\[1ex]
&\quad=|m|\,e^{-\sum_{i,j}^{n-1} (g^{-1})_{ij}\bs{p}_i\cdot \bs{p}_j },
\end{align}
where in the last step  we used the Levi-Civita identity $(n-1)!\,\delta^{[k_1}_{j_1}\ldots \delta^{k_{n-1}]}_{j_{n-1}}=\ep^{j_1\ldots j_{n-1}}\ep_{k_1\ldots k_{n-1}}$ to generate a product of determinants  $|g^{-1}| |G|$, with the $|G|$ then cancelling.
Referring back to \eqref{Schwrep}, since $\mathcal{U}^{-d/2}=|g|^{-d/2}$ we see the action of $\mathcal{S}$ is thus indeed to raise $d\rightarrow d+2$. 

Through similar manipulations, we find
\begin{empheq}[box=\nicebox]{align} \label{Sindef}
\mathcal{S}_{in}^{++} = (-1)^{i+n}\, \frac{(n-1)!}{|G|}\, p_{1}^{[\mu_1}\ldots p_{n-1}^{\mu_{n-1]}} p_{n}^{\mu_{i}}
\prod_{k\neq i}^{n-1}
\frac{\partial}{\partial p_k^{\mu_k}}.
\end{empheq}
Relative to \eqref{Sdef2},  the derivative  $\partial/\partial p_i^{\mu_i}$  has been replaced by the dependent momentum $p_n^{\mu_{i}} =- \sum_{j_i=1}^{n-1}p_{j_i}^{\mu_i}$ positioned to the left of all derivatives.  This leads to 
\begin{align}\label{Sijcalc}
&\mathcal{S}_{in}^{++}\,\Big( e^{-\sum_{i,j}^{n-1} (g^{-1})_{ij}\bs{p}_i\cdot \bs{p}_j }\Big) 
 \nn\\&
=\frac {(-1)^{i+n}(n-1)!}{|G|}\, p_1^{[\mu_1}\ldots p_{n-1}^{\mu_{n-1}]} \nn\\&{\times}
\Big({-}2\sum_{j_1}^{n-1}(g^{-1})_{1{j_1}}p_{j_1}^{\mu_1}\Big)\ldots\Big({-}2\sum_{j_{i}}^{n-1} p_{j_{i}}^{\mu_i}\Big)\ldots
\Big({-}2\sum_{j_{n-1}}^{n-1}(g^{-1})_{n-1,{j_{n-1}}}p_{j_{n-1}}^{\mu_{n-1}}\Big)e^{-\sum_{i,j}^{n-1} (g^{-1})_{ij}\bs{p}_i\cdot \bs{p}_j }\nn\\ &
=(-1)^{i}\,2^{n-2}\sum_{j_i=1}^{n-1}\frac{\partial |g^{-1}|}{\partial (g^{-1})_{i j_{i}}}\,e^{-\sum_{i,j}^{n-1} (g^{-1})_{ij}\bs{p}_i\cdot \bs{p}_j },
\end{align}
since, relative to our previous calculation, the matrix element $(g^{-1})_{ij_i}$ is missing in the product on the middle line.  Instead of obtaining the full determinant $|g^{-1}|$, we then get the derivative of this with respect to the missing element.
As in \eqref{vfromginvn}, we can now
rewrite
\[
\sum_{j_i=1}^{n-1}\frac{\partial |g^{-1}|}{\partial (g^{-1})_{i j_{i}}} = \sum_{j_i=1}^{n-1}(\mathrm{adj}\, g^{-1})_{ij_i} = \sum_{j_i=1}^{n-1}g_{i j_i} |g^{-1}|=v_{in} |g|^{-1}= (-1)^{i}\,2^{2-n}|m^{(i,n)}|
\]
using \eqref{dmdstominnors} in the last step.
The action of $\mathcal{S}_{in}^{++}$ in \eqref{Sindef} on the exponential is thus to pull down a factor of $v_{ij} |g|^{-1}$.  From the representation \eqref{Schwrep}, this has precisely the required action of sending $\alpha_{ij}\rightarrow \alpha_{ij}-1$ and $d\rightarrow d+2$.

Finally, since the choice of dependent momentum is  immaterial,  \eqref{Sindef} generalises to 
\begin{empheq}[box=\nicebox]{align} \label{Sijdef2}
\mathcal{S}_{ij}^{++} =(-1)^{i+j}\, \frac{(n-1)!}{|G|}\, p_{1}^{[\mu_1}\ldots \hat{p}_{j}^{\hat{\mu}_j} \ldots p_{n}^{\mu_{n]}} p_{j}^{\mu_{i}}
\prod_{k\neq i,j}^{n}
\frac{\partial}{\partial p_k^{\mu_k}}
\end{empheq}
where the hats $\hat{p}_{j}^{\hat{\mu}_j}$ indicates that this factor and index are omitted in the antisymmetrised product, and we take $p_j^{\mu_j} = -\sum_{k_j\neq j}^n p_{k_j}^{\mu_j}$ as the dependent momentum. 
In principle these last few derivations allow use of the $s_{ij}$ variables to be avoided entirely, although in practice the form of the operators 
\eqref{Sdef2} and \eqref{Sijdef2} would be hard to anticipate.

\subsection{Further shift operators from shadow conjugation}
\label{shadowsec}

Additional $d\rightarrow d+2$ shift operators can now be constructed -- at no expense -- by conjugating  $\mathcal{S}_{ij}^{++}$ and $\mathcal{S}$ by a pair of shadow transforms.  This idea was discussed recently  for  $d$-preserving weight-shifting operators  in 
\cite{Bzowski:2022rlz}.  

In momentum space, the shadow transform $\Delta_i\rightarrow d - \Delta_i$ (leaving $d$ invariant) simply corresponds to multiplying by $p_i^{d-2\Delta_i}$.  
First, notice that attempting to conjugate $\mathcal{S}_{ij}^{++}$ by shadow transforms on either of  $\Delta_i$ or $\Delta_j$ has no effect:
for example, the action of the operator $p_i^{2\Delta_i-d} \mathcal{S}_{ij}^{++} p_i^{d-2\Delta_i}$ corresponds to the successive parameter shifts
\begin{align}
(\Delta_i,\Delta_j,d) &\xrightarrow{p_i^{d-2\Delta_i}} (d-\Delta_i,\Delta_j,d)\nn\\& \xrightarrow{\mathcal{S}_{ij}^{++}} (d-\Delta_i+1,\Delta_j+1,d+2)\nn\\&
\xrightarrow{p_i^{(d+2)-2(d-\Delta_1+1)}\,=\,p_i^{2\Delta_i-d}} 
((d+2)-(d-\Delta_i+1), \Delta_j+1,d+2) \nn\\
&\qquad\qquad \qquad\qquad \qquad \quad=(\Delta_i+1,\Delta_j+1,d+2)
\end{align}
which is equivalent to the action of $\mathcal{S}_{ij}^{++}$ alone.  Further computations confirm that the shadow transform on $\Delta_i$ or $\Delta_j$ commutes with $\mathcal{S}_{ij}^{++}$.

However, we {\it do} obtain new operators if we shadow conjugate  $\mathcal{S}_{ij}^{++}$ on any index $k\neq i,j$.  For example, the action of 
\[ \label{simplexRop}
p_k^{2\Delta_k+2-d}\mathcal{S}_{ij}^{++} p_k^{d-2\Delta_k}
\] 
corresponds to the successive parameter shifts
\begin{align}\label{simplexRopshifts}
(\Delta_i,\Delta_j,\Delta_k,d) &\xrightarrow{p_k^{d-2\Delta_k}} (\Delta_i,\Delta_j,d-\Delta_k,d)\nn\\& \xrightarrow{\mathcal{S}_{ij}^{++}} (\Delta_i+1,\Delta_j+1,d-\Delta_k,d+2)\nn\\&
\xrightarrow{p_k^{(d+2)-2(d-\Delta_k)}\,=\,p_k^{2\Delta_k+2-d}} 
(\Delta_i+1, \Delta_j+1,\Delta_k+2,d+2).
\end{align}
Thus, in addition  to the shifts produced by $\mathcal{S}_{ij}^{++}$ alone, we have also shifted $\Delta_k$ up by two.
Shadow conjugating on further variables has the same effect, for example, 
\[
p_k^{2\Delta_k+2-d}p_l^{2\Delta_l+2-d}\mathcal{S}_{ij}^{++} p_k^{d-2\Delta_k}p_l^{d-2\Delta_l}
\] 
for  any $(k,l)\neq (i,j)$  sends $(\Delta_i,\Delta_j,\Delta_k,\Delta_l,d)\rightarrow (\Delta_i+1,\Delta_j+1,\Delta_k+2,\Delta_l+2,d+2)$. 

We can also apply similar considerations to $\mathcal{S}$.  The action of
\[
p_i^{2\Delta_i+2-d} \mathcal{S} p_i^{d-2\Delta_i}
\]
corresponds to the shifts
\begin{align}
(\Delta_i,d) \xrightarrow{p_i^{d-2\Delta_i}} (d-\Delta_i,d)
\xrightarrow{\mathcal{S}} (d-\Delta_i, d+2)
\xrightarrow{p_i^{(d+2)-2(d-\Delta_i)}=p_i^{2\Delta_i+2-d}} (\Delta_i+2, d+2).
\end{align}
Shadow conjugating on further momenta $p_k$ leads similarly to shifting $\Delta_k\rightarrow\Delta_k+2$. 

With all these operators obtained by shadow conjugation, notice  we can always obtain an equivalent  differential operator with purely polynomial coefficients ({\it i.e.,} an operator in the Weyl algebra) by commuting the inner $p_k^{d-2\Delta_k}$ shadow factors through the differential operator $\mathcal{S}$ or $\mathcal{S}_{ij}^{++}$, whereupon all non-integer powers cancel with those from the outer shadow transform.

\subsection{Examples at three and four points} 
\label{3n4ptShifts}

To illustrate the general discussion in the two preceding subsections, let us now compute the explicit form of  these $d\rightarrow d+2$ shift operators for $3$- and $4$-point functions. 

\subsubsection{3-point shift operators}

For the 3-point function, it is convenient to use the three squared momentum magnitudes as variables.  Defining
\[
P_i = p_i^2, \qquad D_i = \frac{\partial}{\partial P_i}=\frac{1}{2p_i}\frac{\partial}{\partial p_i}, \qquad i=1,2,3
\]
via momentum conservation we have
\[\label{3ptchain}
P_i = -\sum_{j\neq i}^3 \bs{p}_i\cdot\bs{p}_j = \sum_{j\neq i}^3 V_{ij},\qquad \frac{\partial}{\partial V_{ij}} =\sum_{k=1}^3 \frac{\partial P_k}{\partial V_{ij}}\frac{\partial}{\partial P_k} =  D_i+D_j
\]
From \eqref{Sdef}, writing $D_iD_j=D_{ij}$ for short, we then find
\begin{align}
&\qquad \qquad \mathcal{S} = -4(D_{12}+D_{23}+D_{13}), \nn\\ 
&\mathcal{S}_{12}^{++} = -2D_3, \quad \mathcal{S}_{23}^{++} = -2D_1, \quad \mathcal{S}_{13}^{++} = 2D_2.
\end{align}
The various signs on the second line reflect our choice to use the Cayley minors in \eqref{Cay2} and \eqref{Sdef}: had we used instead the cofactors or $\partial|m|/\partial s_{ij}$ as per \eqref{dmdstominnors} then all signs would be the same. 
Generally, any overall coefficient in $\mathcal{S}$ or the  $\mathcal{S}_{ij}^{++}$ can be eliminated by rescaling the corresponding prefactor in the definition of the simplex integral. 

As noted in the introduction,   these 3-point operators (and their shadow conjugates) are already known from the triple-$K$ representation of the 3-point function.  
In \cite{Bzowski:2013sza, Bzowski:2015yxv}, the Bessel shift operators
\[\label{LRdef}
\mathcal{L}_i = -\frac{1}{p_i}\frac{\partial}{\partial p_i}, \qquad
\mathcal{R}_i = 2\beta_i - p_i\frac{\partial}{\partial p_i} = p_i^{2\beta_i+2}\,\mathcal{L}_i \,p_i^{-2\beta_i},\qquad \beta_i=\Delta_i-\frac{d}{2}
\]
where shown to act on the 3-point function by sending
\begin{align}\label{LRaction}
\mathcal{L}_i: \quad\beta_i\rightarrow \beta_i -1,\quad d\rightarrow d+2, \qquad
\mathcal{R}_i:\quad \beta_i\rightarrow\beta_i+1,\quad d\rightarrow d+2,
\end{align}
or equivalently, 
\begin{align}
&\mathcal{L}_1:\quad  (d,\Delta_1,\Delta_2,\Delta_3)\rightarrow (d+2,\Delta_1,\Delta_2+1,\Delta_3+1), \\
&\mathcal{R}_1:\quad (d,\Delta_1,\Delta_2,\Delta_3)\rightarrow (d+2,\Delta_1+2,\Delta_2+1,\Delta_3+1), 
\end{align}
and similarly under permutations.  This is consistent with our analysis here, since 
\[
(\mathcal{L}_1,\mathcal{L}_2,\mathcal{L}_3)=
(\mathcal{S}_{23}^{++},-\mathcal{S}_{13}^{++},\mathcal{S}_{12}^{++})
\]
and $\mathcal{S}_{ij}^{++}$ augments $\Delta_i$ and $\Delta_j$ by one and $d$ by two.  The $\mathcal{R}_i$ operators are then their shadow conjugates as defined in \eqref{simplexRop}, producing the expected shifts \eqref{simplexRopshifts}.
Finally,
\[
\mathcal{S} = -\mathcal{L}_1\mathcal{L}_2-\mathcal{L}_2\mathcal{L}_3-\mathcal{L}_3\mathcal{L}_1
\]
does not appear explicitly in \cite{Bzowski:2015yxv}, but can be derived as follows.  Writing the 3-point function as the triple-$K$ integral $I_{d/2-1,\{\beta_1,\beta_2,\beta_3\}}$, from \eqref{LRaction} we have
\begin{align}
-\mathcal{S}I_{d/2-1,\{\beta_1,\beta_2,\beta_3\}}& =
I_{d/2+1,\{\beta_1-1,\beta_2-1,\beta_3\}}+
I_{d/2+1,\{\beta_1,\beta_2-1,\beta_3-1\}}+
I_{d/2+1,\{\beta_1-1,\beta_2,\beta_3-1\}}\nn\\[1ex]&=
(\mathcal{R}_1+\mathcal{R}_2+\mathcal{R}_3)I_{d/2,\{\beta_1-1,\beta_2-1,\beta_3-1\}}\nn\\&
=\Big(\frac{d}{2}+\beta_t+4\Big)I_{d/2,\{\beta_1-1,\beta_2-1,\beta_3-1\}}
\end{align}
where the final line follows by eliminating the sum of $\mathcal{R}_i$ operators using the dilatation Ward identity.  The effect of $\mathcal{S}$ is thus to increase $d\rightarrow d+2$ and all $\beta_i\rightarrow \beta_i-1$.  All dimensions $\Delta_i=\beta_i+d/2$ are then preserved, consistent with \eqref{Sijshifts}.

\subsubsection{4-point shift operators}

The 3-point calculations above provide a first consistency check, but to obtain genuinely new shift operators  we now turn to the 4-point function.

To write our results, we introduce the  Mandelstam variables,
\[\label{4ptMands}
P_I = \{p_1^2,p_2^2,p_3^2,p_4^2,s^2,t^2\}, \qquad I=1,\ldots, 6
\]
where $s^2=(\bs{p}_1+\bs{p}_2)^2$ and $t^2=(\bs{p}_2+\bs{p}_3)^2$, and define the derivative operators 
\[
D_I = \frac{\partial}{\partial P_I}, \qquad D_{IJ} = D_I D_J, \qquad D_{IJK} = D_I D_J D_K.
\]
Defining $\mathcal{S}_{ij}^{++} = 4(-1)^{i+j}S_{ij}^{++}$ and $\mathcal{S} = -8 S$ to suppress trivial numerical factors, from \eqref{Sdef} and \eqref{m4}, using the chain rule analogous to \eqref{3ptchain},  we obtain the operators
\begin{align}\label{4ptSij}
&S_{12}^{++} = D_{34}+D_{45}+D_{35}+D_{56}, \nn\\
&S_{13}^{++} = D_{24} - D_{56},\nn\\
&S_{14}^{++} = D_{23}+D_{26}+D_{36}+D_{56},\nn\\
&S_{23}^{++} = D_{14}+D_{16}+D_{46}+D_{56},\nn\\
&S_{24}^{++} = D_{13} - D_{56},\nn\\
&S_{34}^{++} = D_{12}+D_{15}+D_{25}+D_{56},
\end{align}
and 
\begin{align}
S& =D_{456}+D_{356}+D_{346}+D_{256}+D_{246}+D_{245}+D_{235}+D_{234}\nn\\ & \quad +D_{156}+D_{145}+D_{136}+D_{135}+D_{134}+D_{126}+D_{124}+D_{123}.
\end{align}
As per \eqref{Sijshifts}, the $S_{ij}^{++}$ increase $\Delta_i$ and $\Delta_j$ by one and $d$ by two, while $S$ increases $d$ by two.

Following section \ref{shadowsec}, we can obtain further shift operators by shadow conjugation. 
As noted earlier, shadow conjugating each $S_{ij}^{++}$ on either of the $(i,j)$ indices has no effect: from \eqref{4ptSij},  $S_{ij}^{++}$ contains neither $D_i$ or $D_j$ hence these shadow factors commute through the operator.
Instead, we must shadow conjugate each $S_{ij}^{++}$ with respect to indices other than $(i,j)$. 
At four points, once a pair of insertions $(i,j)$ is specified, the remaining set also form a pair $(k,l)\neq (i,j)$. Shadow conjugating each $S_{ij}^{++}$ on the opposite pair $(k,l)$ then  defines
\[
\bar{S}_{ij}^{++}=p_k^{2(\b_k+1)}p_l^{2(\b_l+1)}S_{ij}^{++}p_k^{-2\b_k}p_l^{-2\b_l}, \qquad (k,l)\neq (i,j)
\]
where $\beta_i=\Delta_i-d/2$.  Expressed in terms of the variables \eqref{4ptMands}, we find
\begin{align}
&\bar{S}_{12}^{++}=\b_3\b_4-\b_4P_3D_3-\b_3P_4D_4-(\b_3P_4+\b_4P_3)D_5+P_3P_4S_{12}^{++}, \nn\\
&\bar{S}_{13}^{++}=\b_2\b_4-\b_4P_2D_2-\b_2P_4D_4+P_2P_4S_{13}^{++}, \nn\\
&\bar{S}_{14}^{++}=\b_2\b_3-\b_3P_2D_2-\b_2P_3D_3-(\b_2P_3+\b_3P_2)D_6+P_2P_3S_{14}^{++},\nn\\
&\bar{S}_{23}^{++}=\b_1\b_4-\b_4P_1D_1-\b_1P_4D_4-(\b_1P_4+\b_4P_1)D_6+P_1P_4S_{23}^{++},\nn\\
&\bar{S}_{24}^{++}=\b_1\b_3-\b_3P_1D_1-\b_1P_3D_3+P_1P_3S_{24}^{++},\nn\\
&\bar{S}_{34}^{++}=\b_1\b_2-\b_2P_1D_1-\b_1P_2D_2-(\b_1P_2+\b_2P_1)D_5+P_1P_2S_{34}^{++}.
\end{align}
The action of each operator $\bar{S}_{ij}^{++}$ is to shift $d\rightarrow d+2,~\Delta_{i,j}\rightarrow \Delta_{i,j}+1$ and $\Delta_{k,l}\rightarrow \Delta_{k,l}+2$.  This leaves $\beta_i$ and $\beta_j$ invariant while raising $\beta_k$ and $\beta_l$ by one. 
Heuristically, these $\bar{S}_{ij}^{++}$ are then the 4-point generalisation of the 3-point $\mathcal{R}_i$ operators in \eqref{LRdef}.  Likewise, the $S_{ij}^{++}$ in \eqref{4ptSij} 
leave $\beta_i$ and $\beta_j$ invariant but lower $\beta_{k}$ and $\beta_l$ by one, and represent the 4-point generalisation of the 3-point $\mathcal{L}_i$ operators.

Besides shadow conjugating $S_{ij}^{++}$ with respect to the pair $(k,l)$, one can of course also conjugate with respect to only a single index $k$ to find operators sending $d\rightarrow d+2$, $\Delta_{i,j}\rightarrow \Delta_{i,j}+1$ and $\Delta_k\rightarrow\Delta_k+2$ only.  One can also apply the shadow conjugation procedure to the $d\rightarrow d+2$ operator $S$.  All these operators can be evaluated similarly to the $\bar{S}_{ij}^{++}$ above and we will not write them explicitly.
One case of particular interest, however, corresponds to acting with $\bar{S}_{ij}^{++}$ followed by $S_{ij}^{++}$, which produces an overall shift of $d\rightarrow d+4$ while increasing all operator dimensions by two.  The same shift is produced when acting with these operators in the opposite order (remembering to shift $\beta_{k,l}\rightarrow \beta_{k,l}-1$ in $\bar{S}_{ij}^{++}$  to account for the prior action of $S_{ij}^{++}$).  By subtracting, we then obtain a shift operator of only {\it second} order in derivatives, rather than fourth.  
For example,
\[
\bar{S}_{24}^{++}\Big|_{\beta_1-1,\beta_3-1} S_{24}^{++} - S_{24}^{++} \bar{S}_{24}^{++}\Big|_{\beta_1,\beta_3} = (\beta_1+\beta_3)D_{56}
\]
and so  $D_{56}$ shifts $d\rightarrow d+4$ while sending all $\Delta_i\rightarrow\Delta_i+2$ and preserving the $\beta_i$.

Finally, let us emphasise that the action of all these shift operators  is  general and not in any way tied to the simplex representation: {\it any} solution of the 4-point conformal Ward identities is mapped to an appropriately shifted solution.\footnote{Up to a technical caveat (common to all shift operators) that where divergences occur, one must work in a suitable dimensional regularisation scheme.  In some cases the shift operator then only yields the leading divergences of the shifted correlator, see the discussion in \cite{Bzowski:2022rlz}.} 
We have confirmed this explicitly by computing all the relevant intertwining relations between the shift operators in this section and the conformal Ward identities, whose form in Mandelstam variables can be found in {\it e.g.,} \cite{Bzowski:2022rlz, Arkani-Hamed:2018kmz}.  Thus, for example,
\[\label{intertwiner1}
\mathcal{K}(\Delta_1+1,\Delta_2+1,\Delta_3,\Delta_4,d+2)S_{12}^{++} = S_{12}^{++}\mathcal{K}(\Delta_1,\Delta_2,\Delta_3,\Delta_4,d)
\]
where $\mathcal{K}(\{\Delta_i\},d)$ represents schematically any of the special conformal or dilatation Ward identities with the operator and spacetime dimensions as indicated.  Applying this relation to any CFT correlator with dimensions $(\{\Delta_i\},d)$, the right-hand side vanishes and the left-hand side then indicates that the action of $S_{12}^{++}$ produces a solution of the shifted Ward identities.
Intertwining relations such as these\footnote{More generally,  the right-hand side of \eqref{intertwiner1} could feature {\it any} operator in the left ideal of the conformal Ward identities, since all that matters is that it vanishes when acting on a solution with dimensions $(\{\Delta_i\},d)$.}  allow the shift action of operators to be established independently of any integral representation for the correlator.

\subsection[{\texorpdfstring{Operators preserving $d$}{Operators preserving d}}]{\texorpdfstring{Operators preserving \boldmath{$d$}}{Operators preserving d}}
\label{sec:Wops}

A different class of weight-shifting operators that preserve the spacetime dimension $d$ while shifting the $\Delta_i$ was identified in \cite{Karateev:2017jgd}.
In momentum space, these operators have been applied to de Sitter correlators in \cite{Arkani-Hamed:2018kmz, Baumann:2019oyu}.
With the aid of shadow conjugation, we can write them in the compact form  \cite{Bzowski:2022rlz}
\begin{align}
\label{Wops}
\mathcal{W}^{--}_{ij} &=\frac{1}{2}\, \Big(\frac{\partial}{\partial p_i^\mu}-\frac{\partial}{\partial p_j^\mu}\Big)\Big(\frac{\partial}{\partial p_{i\mu}}-\frac{\partial}{\partial p_{j \mu}}\Big)
\nn\\
\mathcal{W}^{+-}_{ij}&= p_i^{2(\beta_i+1)}\mathcal{W}^{--}_{ij} p_i^{-2\beta_i}\nn\\
\mathcal{W}^{-+}_{ij} &= p_j^{2(\beta_j+1)}\mathcal{W}^{--}_{ij} p_j^{-2\beta_j} \nn\\
\mathcal{W}_{ij}^{++}  &=
p_i^{2(\beta_i+1)}p_j^{2(\beta_j+1)} \mathcal{W}_{ij}^{--}p_i^{-2\beta_i}p_j^{-2\beta_j},
\end{align}
where $\beta_i=\Delta_i-d/2$ and $1\le i<j\le n-1$ so $\bs{p}_n$ is taken as the dependent momentum.  
Their action is to shift
\[\label{Waction}
\mathcal{W}^{\sigma_i\sigma_j}_{ij}:\quad \Delta_i\rightarrow\Delta_i+\sigma_i,\quad \Delta_j\rightarrow\Delta_j+\sigma_j,\quad d\rightarrow d, \quad \{\sigma_i,\sigma_j\}\in \pm1.
\]

In this section, our goal is to understand the action of the simplest of these operators, $\mathcal{W}_{ij}^{--}$, from the simplex perspective.  The action of the others then follows via shadow conjugation, or else can be shown explicitly: for example, we analyse  $\mathcal{W}_{ij}^{-+}$ in section \ref{sec:Wmp}.

We begin by writing the Schwinger exponential \eqref{Kirchhoff1} in the form
\[\label{SchwExpRes}
-\mathrm{tr}(g^{-1}\cdot G)  
=\sum_{k<l}^n s_{kl}\,\bs{p}_k\cdot\bs{p}_l
=-\sum_{k<l}^{n-1} (s_{kn}+s_{ln}-s_{kl})(\bs{p}_k\cdot\bs{p}_l) -\sum_{k}^{n-1}s_{kn}\,p_k^2.
\]
As only the independent momenta feature in this last expression, the action of $\mathcal{W}_{ij}^{--}$ on the Schwinger exponential can  be rewritten as a differential operator in the $s_{kl}$.  We will do this in several steps.  First, notice that 
\begin{align}
\frac{\partial}{\partial p_i^\mu} e^{-\mathrm{tr}(g^{-1}\cdot G) } &=-\Big( 2s_{in}p_i^\mu
+\sum_{k\neq i}^{n-1}(s_{in}+s_{kn}-s_{ik})p_k^\mu \Big) e^{-\mathrm{tr}(g^{-1}\cdot G) } \nn\\&
=-\sum_{k}^{n-1}(s_{in}+s_{kn}-s_{ik})\,p_k^\mu\, e^{-\mathrm{tr}(g^{-1}\cdot G) },
\end{align}
where in the second line  $s_{ik}$ vanishes for $i=k$.   This gives
\begin{align}
\Big(\frac{\partial}{\partial p_i^\mu} -\frac{\partial}{\partial p_j^\mu} \Big)e^{-\mathrm{tr}(g^{-1}\cdot G) } &=
\sum_{k}^{n-1}(s_{ik}-s_{jk}-s_{in}+s_{jn})\,p_k^\mu \,e^{-\mathrm{tr}(g^{-1}\cdot G) },
\end{align}
and hence
\begin{align}
&\mathcal{W}_{ij}^{--}e^{-\mathrm{tr}(g^{-1}\cdot G) }\\  &\quad =
\Big(-d s_{ij} + \frac{1}{2}\sum_{k,l}^{n-1}(s_{ik}-s_{jk}-s_{in}+s_{jn})(s_{il}-s_{jl}-s_{in}+s_{jn})\,\bs{p}_k\cdot\bs{p}_l\Big)\,e^{-\mathrm{tr}(g^{-1}\cdot G) }.\nn
\end{align}
To rewrite these momentum dot products as derivatives with respect to the $s_{kl}$, we now rearrange this sum as follows.  Using momentum conservation $p_k^2 =-\sum_{l\neq k}^n\bs{p}_k\cdot \bs{p}_l$, for any generic coefficient $A_k$ such that $A_n=0$, we have
\begin{align}\label{Atrick}
\sum_{k,l}^{n-1}A_k A_l \,\bs{p}_k\cdot\bs{p}_l &= \sum_{\substack{k,l\\k\neq l}}^{n-1}A_k A_l \,\bs{p}_k\cdot\bs{p}_l
+\sum_{k}^{n-1}A_k^2\, p_k^2 \nn\\
& = \sum_{\substack{k,l\\k\neq l}}^{n-1}A_k (A_l-A_k) \,\bs{p}_k\cdot\bs{p}_l -\sum_{k}^{n-1}A_k^2 \,\bs{p}_k\cdot\bs{p}_n
\nn\\
& = -\frac{1}{2}\sum_{\substack{k,l\\k\neq l}}^{n-1} (A_l-A_k)^2 \,\bs{p}_k\cdot\bs{p}_l -\sum_{k}^{n-1}A_k^2 \,\bs{p}_k\cdot\bs{p}_n
\nn\\
& = -\sum_{k<l}^{n} (A_l-A_k)^2 \,\bs{p}_k\cdot\bs{p}_l 
\end{align}
where in the final line the sum runs up to $n$.
Setting $A_k = s_{ik}-s_{jk}-s_{in}+s_{jn}$, we find
\begin{align}\label{Wmmid}
\mathcal{W}_{ij}^{--}e^{-\mathrm{tr}(g^{-1}\cdot G) } &=
\Big(-d s_{ij} - \frac{1}{2}\sum_{k<l}^{n} (s_{ik}-s_{jk}-s_{il}+s_{jl})^2\,\bs{p}_k\cdot\bs{p}_l
\Big)\,e^{-\mathrm{tr}(g^{-1}\cdot G) }\nn\\
&
=
\Big(-d s_{ij} - \frac{1}{2}\sum_{k<l}^{n} (s_{ik}-s_{jk}-s_{il}+s_{jl})^2\,\partial_{s_{kl}}
\Big)\,e^{-\mathrm{tr}(g^{-1}\cdot G) }\nn\\&
=(-d s_{ij} + 2\partial_{v_{ij}})\,e^{-\mathrm{tr}(g^{-1}\cdot G) }\nn\\[1ex]
& = 2\, |g|^{d/2} \,\partial_{v_{ij}} \,\Big(|g|^{-d/2}\,e^{-\mathrm{tr}(g^{-1}\cdot G) }\Big).
\end{align}
In the second line here, we exchanged $\bs{p}_k\cdot\bs{p}_l$ for $\partial_{s_{kl}}$ using the first expression in \eqref{SchwExpRes}.
The change of variables from $\partial_{s_{kl}}$ to $\partial_{v_{ij}}$ in the third line then comes from the Jacobian evaluated in appendix \ref{sec:Jacobimatrixels}, and in the final line we used \eqref{sdef}.

The action of $\mathcal{W}_{ij}^{--}$ on the full simplex integral \eqref{Schwrep} now follows.  First, the outer factor of $|g|^{d/2}$ in \eqref{Wmmid} cancels with the factor $\mathcal{U}^{-d/2}=|g|^{-d/2}$ in \eqref{Schwrep}.  Integrating by parts with respect to $v_{ij}$, assuming the boundary terms vanish,\footnote{For the upper limit this is automatic for momentum configurations with non-vanishing Gram determinant thanks to the decaying exponential.  The lower limit vanishes provided $\alpha_{ij}<0$.} 
the derivative then acts on the prefactors  as
\[
-2\partial_{v_{ij}}\Big(\prod_{k<l}^n v_{kl}^{-\alpha_{kl}-1} f(\hat{\bs{v}}) \Big)= v_{ij}^{-1}\prod_{k<l}^n v_{kl}^{-\alpha_{kl}-1} \tilde{f}(\hat{\bs{v}}).  
\]
Here, the terms coming from $\partial_{v_{ij}}$ hitting the cross-ratios \eqref{vcrossratio} inside the arbitrary function $f(\hat{\bs{v}})$, as well as those from hitting $v_{ij}^{-\alpha_{ij}-1}$, have been repackaged  in the form $v_{ij}^{-1}\tilde{f}(\hat{\bs{v}})$ for some new function of cross-ratios  $\tilde{f}(\hat{\bs{v}})$.  Thus, overall, we find 
\begin{align}
&\mathcal{W}_{ij}^{--} \Big(\prod_{k<l}^n \int_0^\infty \mathrm{d}v_{kl}\, v_{kl}^{-\alpha_{kl}-1}\Big) \,f(\hat{\bs{v}}) |g|^{-d/2} e^{-\mathrm{tr}(g^{-1}\cdot G)} \nn\\
&\qquad \quad =\Big(
\prod_{k<l}^n \int_0^\infty \mathrm{d}v_{kl}\,  v_{kl}^{-\alpha_{kl}-1} \Big)\,v_{ij}^{-1}\tilde{f}(\hat{\bs{v}})|g|^{-d/2} e^{-\mathrm{tr}(g^{-1}\cdot G)}.
\end{align}
The action of $\mathcal{W}_{ij}^{--}$ on the simplex is therefore to send $\alpha_{ij}\rightarrow \alpha_{ij}+1$, up to changes of the arbitrary function.  The latter is of no account as far as mapping one solution of the conformal Ward identities to another is concerned.\footnote{An exception is if  $\mathcal{W}_{ij}^{--}$ maps us from a finite correlator to a singular one, corresponding to a solution of the conditions $d+\sum_{i=1}^n \sigma_i(\Delta_i-d/2)=-2k$ for some non-negative integer $k$ and a choice of signs $\{\sigma_i\}\in \pm 1$, see \cite{Bzowski:2019kwd}.  In such cases, the arbitrary function $\tilde{f}(\hat{\bs{v}})$ vanishes.  In dimensional regularisation, this zero then cancels the pole coming from the divergent correlator such that the result is finite, see \cite{Bzowski:2022rlz}.}    
From \eqref{alphaijdef}, we now confirm that sending $\alpha_{ij}\rightarrow \alpha_{ij}+1$ while keeping the remaining $\alpha_{kl}$ fixed is equivalent to sending
$\Delta_i\rightarrow \Delta_i-1$ and $\Delta_j\rightarrow \Delta_j-1$ while preserving $d$, in perfect agreement with \eqref{Waction}.

\section{Verifying the conformal Ward identities}
\label{sec:CWI}

In this section, we 
prove that the  parametric representation of the simplex integral \eqref{Schwrep} satisfies the conformal Ward identities for any arbitrary function of cross-ratios.
The corresponding result for the vectorial simplex integral \eqref{simplex} was established in  \cite{Bzowski:2019kwd, Bzowski:2020kfw}.
Working purely in momentum space,
our  approach is 
to show that the action of the Ward identities on the simplex integral reduces to a total derivative.  With a degree of hindsight, the structure of this total derivative, obtained in  \eqref{CWItotalderiv}, 
can also be understood from somewhat simpler position-space arguments.  We will return to these in section \ref{sec:posnspacecwis}.

As the dilatation Ward identity can be verified by power counting, we focus on the special conformal Ward identities
\[\label{scwi}
0 =  \sum_{j=1}^{n-1}\Big(p_j^\mu \frac{\partial}{\partial p_j^\nu}\frac{\partial}{\partial p_j^\nu}-2p_j^\nu \frac{\partial}{\partial p_j^\nu}\frac{\partial}{\partial p_j^\mu}+2(\Delta_j-d)\frac{\partial}{\partial p_j^\mu}\Big)\lla \O_1(\bs{p}_1) \ldots \O_n(\bs{p}_n) \rra,
\]
treating $\bs{p}_n$ as the dependent momentum.
As a first step, we rewrite the action of each individual term in \eqref{scwi} on the Schwinger exponential as an equivalent differential operator in $v_{ij}$.   
From \eqref{Kirchhoff1}, we have
\begin{align}
\sum_{j}^{n-1}2(\Delta_j-d)\frac{\partial}{\partial p_j^\mu} \,e^{-\mathrm{tr}\,(g^{-1}\cdot G)} &= \sum_{j}^{n-1} p_j^\mu\Big( -4\sum_{k}^{n-1} (\Delta_k-d)g^{-1}_{jk} \Big)\,e^{-\mathrm{tr}\,(g^{-1}\cdot G)} ,\\
\sum_{j}^{n-1}p_j^\mu \frac{\partial}{\partial p_j^\nu}\frac{\partial}{\partial p_j^\nu}\,e^{-\mathrm{tr}\,(g^{-1}\cdot G)} &=\sum_{j}^{n-1} p_j^\mu\Big(-2d g^{-1}_{jj}+ 4\sum_{k,l}^{n-1}\,g^{-1}_{jk}g^{-1}_{jl} \,\bs{p}_k\cdot\bs{p}_l\Big)\,e^{-\mathrm{tr}\,(g^{-1}\cdot G)}. 
\end{align}
Using \eqref{ginvs} for the inverse metric and the manipulation \eqref{Atrick}, this last expression can be rewritten analogously to \eqref{Wmmid}:
\begin{align}
\sum_{j}^{n-1}p_j^\mu \frac{\partial}{\partial p_j^\nu}\frac{\partial}{\partial p_j^\nu}\,e^{-\mathrm{tr}\,(g^{-1}\cdot G)} &=\sum_{j}^{n-1} p_j^\mu\Big(-2d s_{jn}-\sum_{k<l}^{n}\,(s_{jk}-s_{kn}-s_{jl}+s_{ln})^2  \,\bs{p}_k\cdot\bs{p}_l\Big)\,e^{-\mathrm{tr}\,(g^{-1}\cdot G)}\nn\\
 &=\sum_{j}^{n-1} p_j^\mu\Big(-2d s_{jn}-\sum_{k<l}^{n}\,(s_{jk}-s_{kn}-s_{jl}+s_{ln})^2 \,\partial_{s_{kl}}\Big)\,e^{-\mathrm{tr}\,(g^{-1}\cdot G)}\nn\\
  & =\sum_{j}^{n-1} p_j^\mu\Big(-2d s_{jn}+4\partial_{v_{jn}}\Big)\,e^{-\mathrm{tr}\,(g^{-1}\cdot G)}. 
\end{align}
Next, we must deal with
\begin{align}
&\sum_{j}^{n-1}\Big(-2p_j^\nu \frac{\partial}{\partial p_j^\nu}\frac{\partial}{\partial p_j^\mu}\Big)\,e^{-\mathrm{tr}\,(g^{-1}\cdot G)} \nn\\&\qquad\qquad =4\sum_{j}^{n-1} p_j^\mu\,\Big( g_{jj}^{-1}-\sum_{k,l}^{n-1}(g^{-1}_{jk}+g^{-1}_{jl})g^{-1}_{kl}\,\bs{p}_k\cdot\bs{p}_l\Big)\,e^{-\mathrm{tr}\,(g^{-1}\cdot G)}\nn\\&
\qquad\qquad 
=4\sum_{j}^{n-1} p_j^\mu\,\Big( g_{jj}^{-1}-2\sum_{k<l}^{n-1}(g^{-1}_{jk}+g^{-1}_{jl})g^{-1}_{kl}\,\bs{p}_k\cdot\bs{p}_l
-2\sum_{k}^{n-1} g^{-1}_{jk}g^{-1}_{kk} p_k^2
\Big)\,e^{-\mathrm{tr}\,(g^{-1}\cdot G)}. \label{gkkpiece1}
\end{align}
Using \eqref{ginvs} and momentum conservation, the $p_k^2$ terms in this final sum can be rewritten 
\begin{align}
-2\sum_{k}^{n-1} g^{-1}_{jk}g^{-1}_{kk} p_k^2 &=
\sum_{k}^{n}s_{kn}(s_{jn}+s_{kn}-s_{jk}) \sum_{l\neq k}^n \bs{p}_k\cdot\bs{p}_l \nn\\
& = \sum_{k<l}^n \Big(s_{kn}(s_{jn}+s_{kn}-s_{jk})+s_{ln}(s_{jn}+s_{ln}-s_{jl})\Big)\bs{p}_k\cdot\bs{p}_l. \label{gkkpiece2}
\end{align}
In the first line here, notice we extended the sum over $k$ to run up to $n$, which is possible since the additional term with $k=n$ vanishes as $s_{nn}=0$.  To get the second line, we then re-expressed the terms for which $k>l$ by swapping $k\leftrightarrow l$. 
For convenience, it is useful to define
\[\label{hatgdef}
\hat{g}^{-1}_{ij} = \frac{1}{2}(s_{in}+s_{jn}-s_{ij}) = \begin{cases}
g^{-1}_{ij}\qquad & i,j\le n-1,\\
0\qquad & i=n \,\,\mathrm{and/or}\,\, j=n,
\end{cases}
\]
effectively extending the $(n-1)\times(n-1)$ matrix $g^{-1}_{ij}$ to an $n\times n$ matrix $\hat{g}^{-1}_{ij}$ by adding a final row and column of zeros. 
This allows us to compactly rewrite \eqref{gkkpiece1} and \eqref{gkkpiece2} as
\begin{align}
&\sum_{j}^{n-1}\Big(-2p_j^\nu \frac{\partial}{\partial p_j^\nu}\frac{\partial}{\partial p_j^\mu}\Big)\,e^{-\mathrm{tr}\,(g^{-1}\cdot G)} \nn\\&\quad
=\sum_{j}^{n-1} p_j^\mu\,\Big( 4\hat{g}_{jj}^{-1}+8\sum_{k<l}^{n}\Big(-(\hat{g}^{-1}_{jk}+\hat{g}^{-1}_{jl})\hat{g}^{-1}_{kl}+\hat{g}_{kk}^{-1}\hat{g}^{-1}_{jk}+\hat{g}^{-1}_{ll}\hat{g}^{-1}_{jl}\Big)\,\partial_{s_{kl}}
\Big)\,e^{-\mathrm{tr}\,(g^{-1}\cdot G)}.
\end{align}
Here, the sum over $l$  for  the $\bs{p}_k\cdot\bs{p}_l$ terms in \eqref{gkkpiece1} has similarly been extended to run up to $n$, noting the additional $l=n$ term vanishes.  We then replaced $\bs{p}_k\cdot\bs{p}_l$  by a derivative with respect to $s_{kl}$ using \eqref{SchwExpRes}.  
The result now simplifies  further upon exchanging 
\[
\partial_{s_{kl}} = \sum_{a<b}^n \frac{\partial v_{ab}}{\partial s_{kl}}\, \partial_{v_{ab}}=  -\sum_{a<b}^n \tilde{g}_{a(k}\tilde{g}_{l)b}\,
\partial_{v_{ab}},
\]
where $\tilde{g}_{ab}$ is the Laplacian matrix \eqref{tildeg} and the Jacobian is evaluated in appendix \ref{sec:Jacobimatrixels}.
 First, we write
\begin{align}
&8\sum_{k<l}^{n}\Big(-(\hat{g}^{-1}_{jk}+\hat{g}^{-1}_{jl})\hat{g}^{-1}_{kl}+\hat{g}_{kk}^{-1}\hat{g}^{-1}_{jk}+\hat{g}^{-1}_{ll}\hat{g}^{-1}_{jl}\Big)\,\partial_{s_{kl}}\nn\\
&\quad =
4\sum_{a<b}^n\sum_{k,l}^{n}\tilde{g}_{ak}\tilde{g}_{lb}\,\Big((\hat{g}^{-1}_{jk}+\hat{g}^{-1}_{jl})\hat{g}^{-1}_{kl}-\hat{g}_{kk}^{-1}\hat{g}^{-1}_{jk}-\hat{g}^{-1}_{ll}\hat{g}^{-1}_{jl}\Big)\,\partial_{v_{ab}} \label{inter1}
\end{align}
where the sum over $k<l$ of $(k,l)$-symmetric terms has been rewritten as half the sum over all $k$ and $l$, noting the terms with $k=l$ explicitly cancel.  
The final two terms now vanish since all row and column sums of the Laplacian matrix $\tilde{g}$ are zero: 
\begin{align}
 \sum_{k,l}^{n}\tilde{g}_{ak}\,\tilde{g}_{lb}\,\hat{g}^{-1}_{kk}\,\hat{g}^{-1}_{jk}&=
  \sum_{k}^{n}\tilde{g}_{ak}\,\hat{g}^{-1}_{kk}\,\hat{g}^{-1}_{jk}
 \sum_{l}^n \,\tilde{g}_{lb} =0,\\
  \sum_{k,l}^{n}\tilde{g}_{ak}\,\tilde{g}_{lb}\,\hat{g}^{-1}_{ll}\,\hat{g}^{-1}_{jl}&=
  \sum_{k}^{n}\tilde{g}_{lb}\,\hat{g}^{-1}_{ll}\,\hat{g}^{-1}_{jl}
 \sum_{k}^n \,\tilde{g}_{ak} =0. 
\end{align}
For the first two terms in \eqref{inter1}, we use the identity
\[
\sum_{k}^n \tilde{g}_{ik}\,\hat{g}^{-1}_{kj}=\delta_{ij}-\delta_{in}, \qquad i,j\le n.
\]
To derive this, note the sum over $k$ restricts to $k\le n-1$ from  \eqref{hatgdef}, then for $i,j\le n-1$ we have $\tilde{g}_{ik}\hat{g}^{-1}_{kj}=g_{ik}g^{-1}_{kj}$.  For $i=n$, $j\le n-1$ we use $\tilde{g}_{nk}\hat{g}^{-1}_{kj}=-\sum_{l}^{n-1}g_{lk}g^{-1}_{kj}$ and for $j=n$ and any $i$ the sum vanishes from \eqref{hatgdef}.  
With the aid of this identity, we then find
\begin{align}
&\sum_{j}^{n-1}\Big(-2p_j^\nu \frac{\partial}{\partial p_j^\nu}\frac{\partial}{\partial p_j^\mu}\Big)\,e^{-\mathrm{tr}\,(g^{-1}\cdot G)} 
\nn\\&\qquad
=4\sum_{j}^{n-1} p_j^\mu\,\Big( s_{jn} -\partial_{v_{jn}}-\sum_{a<b}^{n}(\hat{g}^{-1}_{ja}+\hat{g}^{-1}_{jb})\theta_{v_{ab}}
\Big)\,e^{-\mathrm{tr}\,(g^{-1}\cdot G)} 
\end{align}
where we used $\tilde{g}_{ab}=-v_{ab}$ for $a<b$ to obtain the Euler operator $\theta_{v_{ab}}=v_{ab}\partial_{v_{ab}}$.

Assembling the pieces above, the action of the conformal Ward identity is now
\begin{align}
&\sum_{j=1}^{n-1}\Big(p_j^\mu \frac{\partial}{\partial p_j^\nu}\frac{\partial}{\partial p_j^\nu}-2p_j^\nu \frac{\partial}{\partial p_j^\nu}\frac{\partial}{\partial p_j^\mu}+2(\Delta_j-d)\frac{\partial}{\partial p_j^\mu}\Big)\,e^{-\mathrm{tr}\,(g^{-1}\cdot G)} 
\nn\\&\quad
=4\sum_{j}^{n-1}p_j^\mu\Big(\Big(1-\frac{d}{2}\Big) s_{jn} - \sum_{k}^{n-1}(\Delta_k-d)g^{-1}_{jk}
-\sum_{a<b}^{n}(\hat{g}^{-1}_{ja}+\hat{g}^{-1}_{jb})\theta_{v_{ab}}
\Big)\,e^{-\mathrm{tr}\,(g^{-1}\cdot G)} 
\nn\\&\quad
=4\sum_{j}^{n-1}p_j^\mu\Big(\Big(1-\frac{d}{2}\Big)  s_{jn}  + d\sum_{a}^{n}\hat{g}^{-1}_{ja}
+\sum_{a\neq b}^{n}\hat{g}^{-1}_{ja}(\alpha_{ab}-\theta_{v_{ab}})
\Big)\,e^{-\mathrm{tr}\,(g^{-1}\cdot G)} 
\label{CWIres1}
\end{align}
using \eqref{alphaijdef} in the last line.
Finally, we need two further identities:
\[\label{twoids}
\sum_{a\neq b}^n \theta_{v_{ab}} \hat{g}^{-1}_{ja}  = -s_{jn}, \qquad
 \sum_{a\neq b}^n \hat{g}^{-1}_{ja}v_{ab}s_{ab} = -s_{jn}+2\sum_a^{n}\hat{g}^{-1}_{ja}
\]
To establish the first of these, we write
\begin{align}
\sum_{a\neq b}^n \theta_{v_{ab}} \hat{g}^{-1}_{ja}  
= 
-\sum_{a\neq b}^{n-1} g_{ab}\frac{\partial g_{ja}^{-1}}{\partial v_{ab}}+\sum_{a}^{n-1}v_{an}\frac{\partial g^{-1}_{ja}}{\partial v_{an}}
= 
-\sum_{a\neq b}^{n-1} g_{ab}\frac{\partial g_{ja}^{-1}}{\partial v_{ab}}+\sum_{a,b}^{n-1}g_{ab}\frac{\partial g^{-1}_{ja}}{\partial v_{an}}
\label{thetaid1}
\end{align}
then use the chain rule, which for $i,j,k,l\le n-1$ gives 
\begin{align}
\frac{\partial g^{-1}_{ij}}{\partial v_{kl}} &= -\sum_{a,b}^{n-1}g^{-1}_{i(a}g^{-1}_{b)j}\,\frac{\partial g_{ab}}{\partial v_{kl}}
= (g^{-1}_{ik}-g^{-1}_{il})(g^{-1}_{jl}-g^{-1}_{jk}), \\
\frac{\partial g^{-1}_{ij}}{\partial v_{kn}} &= -\sum_{a,b}^{n-1}g^{-1}_{i(a}g^{-1}_{b)j}\,\frac{\partial g_{ab}}{\partial v_{kn}} = - g^{-1}_{ik}g^{-1}_{kj}.
\end{align}
Inserting these into \eqref{thetaid1}, the sum over $a\neq b$ can be extended to run over all $a,b$ since the term with $a=b$ vanishes.  The only non-cancelling term is then $-g^{-1}_{jj}=-s_{jn}$ as required. 

For the second identity in \eqref{twoids}, we use  \eqref{ginvs} to rewrite
\begin{align}
\sum_{a\neq b}^n \hat{g}^{-1}_{ja}v_{ab}s_{ab} &= -\sum_{a\neq b}^{n-1} g^{-1}_{ja}g_{ab}s_{ab} +\sum_a^{n-1}g^{-1}_{ja}v_{an}s_{an} \nn\\&
=-\sum_{a\neq b}^{n-1} g^{-1}_{ja}g_{ab}(g^{-1}_{aa}+g^{-1}_{bb}-2g^{-1}_{ab})+ \sum_a^{n-1}g^{-1}_{ja}\Big(\sum_b^{n-1}g_{ab}\Big)g^{-1}_{aa}.
\end{align}
The sum over $a\neq b$ can then be extended to run over all $a,b$ as the term with $a=b$ cancels, after which the  first and the last terms cancel and the result follows.

With the aid of the identities \eqref{twoids}, we find that \eqref{CWIres1} becomes
\begin{align}
&\sum_{j=1}^{n-1}\Big(p_j^\mu \frac{\partial}{\partial p_j^\nu}\frac{\partial}{\partial p_j^\nu}-2p_j^\nu \frac{\partial}{\partial p_j^\nu}\frac{\partial}{\partial p_j^\mu}+2(\Delta_j-d)\frac{\partial}{\partial p_j^\mu}\Big)\,e^{-\mathrm{tr}\,(g^{-1}\cdot G)} 
\nn\\&\quad
=-4\sum_{j}^{n-1}p_j^\mu\,
|g|^{d/2}\Omega^{-1}\sum_{a\neq b}^{n}\partial_{v_{ab}}\Big(v_{ab}\,\hat{g}^{-1}_{ja}|g|^{-d/2}\Omega
\,e^{-\mathrm{tr}\,(g^{-1}\cdot G)} \Big)
\end{align}
where 
$
\Omega=\prod_{k<l}^n v_{kl}^{-\alpha_{kl}-1}.
$
Recalling that the simplex representation \eqref{Schwrep} is 
\begin{align}
\lla \O_1(\bs{p}_1) \ldots \O_n(\bs{p}_n) \rra=\Big(\prod_{k<l}^n\int_0^\infty \D v_{kl}\,v_{kl}^{-\alpha_{kl}-1}\Big) f(\hat{\bs{v}}) |g|^{-d/2} \,e^{-\mathrm{tr}\,(g^{-1}\cdot G)}, 
\end{align}
we note that
\[\label{sumofEuleronf}
\sum_{\substack{b\\b\neq a}}^n \theta_{v_{ab}}f(\hat{\bs{v}})=0
\]
since whenever the index $a$ appears in a cross ratio $\hat{v}_{[acde]} = v_{ac}v_{de}/v_{ad}v_{ce}$ it enters with equal weight in the numerator and the denominator producing a cancellation.
Acting with the Ward identity thus yields a total derivative:
\begin{align}
&\sum_{j=1}^{n-1}\Big(p_j^\mu \frac{\partial}{\partial p_j^\nu}\frac{\partial}{\partial p_j^\nu}-2p_j^\nu \frac{\partial}{\partial p_j^\nu}\frac{\partial}{\partial p_j^\mu}+2(\Delta_j-d)\frac{\partial}{\partial p_j^\mu}\Big)\,\lla \O_1(\bs{p}_1) \ldots \O_n(\bs{p}_n) \rra\nn\\&\quad
=-4\sum_{j}^{n-1}p_j^\mu\Big(\prod_{k<l}^n\int_0^\infty \D v_{kl}\Big)  \sum_{a\neq b}^{n}\partial_{v_{ab}}\Big(v_{ab}\,\hat{g}^{-1}_{ja} \,f(\hat{\bs{v}})|g|^{-d/2}\Omega
\,e^{-\mathrm{tr}\,(g^{-1}\cdot G)} \Big).
\label{CWItotalderiv}
\end{align}
The boundary terms vanish under reasonable assumptions: for generic  momentum configurations with non-vanishing Gram determinant, the upper limit is suppressed by the decay of the Schwinger exponential; the lower limit is zero 
provided $v_{ab}^{-\alpha_{ab}}f(\hat{\bs{v}})$ vanishes as $v_{ab}\rightarrow 0$, which is satisfied whenever the simplex representation itself converges.
The simplex integral thus solves the special conformal Ward identity.

\section{Insight from position space}
\label{sec:posnspace}

Thus far, our analysis has been entirely in momentum space.  However, as noted above, the form of the total derivative produced by the action of the special conformal Ward identity in  \eqref{CWItotalderiv} 
can also  be understood through independent position-space arguments.  
We present these in section \ref{sec:posnspacecwis}.  Then, in section \ref{sec:Wmp}, we  show how similar position-space arguments can  be applied to verify the action of $d$-preserving shift operators such as $\mathcal{W}_{12}^{-+}$.

\subsection{The conformal Ward identities}
\label{sec:posnspacecwis}

To Fourier transform the simplex representation \eqref{Schwrep}  to position space, we  compute
\begin{align}
&\<\O(\x_1)\ldots \O(\x_n)\> = \prod_{k}^{n-1}\int\frac{\D^d\bs{p}_k}{(2\pi)^d}\,e^{i\bs{p}_k\cdot \bs{x}_{kn}} \lla \O(\bs{p}_1)\ldots \O(\bs{p}_n)\rra\nn\\&\quad =
\Big(\prod_{i<j}^n\int\D v_{ij}\, v_{ij}^{-\alpha_{ij}-1}\Big)f(\hat{\bs{v}}) |g|^{-d/2} \Big(\prod_{k}^{n-1}\int\frac{\D^d\bs{p}_k}{(2\pi)^d}\Big)\exp\Big(\sum_{k}^{n-1}i\bs{p}_k\cdot \bs{x}_{kn} - \sum_{k,l}^{n-1}g^{-1}_{kl} \,\bs{p}_k\cdot \bs{p}_l\Big)\nn\\
&\quad =
\Big(\prod_{i<j}^n\int\D v_{ij}\, v_{ij}^{-\alpha_{ij}-1}\Big)\tilde{f}(\hat{\bs{v}}) \exp\Big(-\frac{1}{4}\sum_{k,l}^{n-1}g_{kl}\,\bs{x}_{kn}\cdot\bs{x}_{ln}\Big) \label{possp1}
\end{align}
where $\bs{x}_{ij}=\bs{x}_i-\bs{x}_j$, and for the Gaussian integral over momenta we completed the square: 
\begin{align}
&\sum_{k}^{n-1}i\bs{p}_k\cdot \bs{x}_{kn} - \sum_{k,l}^{n-1}g^{-1}_{kl} \,\bs{p}_k\cdot \bs{p}_l \nn\\\qquad  &=-\sum_{k,l}^{n-1}g^{-1}_{kl}(\bs{p}_k-\frac{i}{2}\sum_{a}^{n-1}g_{ka}\,\bs{x}_{an})\cdot(\bs{p}_l-\frac{i}{2}\sum_{b}^{n-1}g_{lb}\,\bs{x}_{bn}) -\frac{1}{4}\sum_{k,l}^{n-1}g_{kl}\,\bs{x}_{kn}\cdot\bs{x}_{ln}.
\end{align}
The numerical factor from the integration can then be re-absorbed into the arbitrary function by setting $(4\pi)^{(1-n)d/2} f(\hat{\bs{v}})=\tilde{f}(\hat{\bs{v}})$.
The exponent in \eqref{possp1} now simplifies to\footnote{Recall the analogous relation  in a resistor network of simplex topology, namely, that the power dissipated is $\sum_{i<j}^n v_{ij}(V_i-V_j)^2 = \sum_{i,j}^n \tilde{g}_{ij}V_i V_j$, where $v_{ij}$ is the conductivity and $V_i$ the voltage at node $i$.}
\begin{align}
-\frac{1}{4}\sum_{k,l}^{n-1}g_{kl}\,\bs{x}_{kn}\cdot\bs{x}_{ln}
=-\frac{1}{4}\sum_{k,l}^{n}\tilde{g}_{kl}\,\bs{x}_{k}\cdot\bs{x}_{l}=-\frac{1}{4}\sum_{i<j}^{n} v_{ij}x_{ij}^2,
\end{align}
and hence the simplex representation in position space is
\begin{align}\label{posnsimplex}
&\<\O(\x_1)\ldots \O(\x_n)\> =
\Big(\prod_{i<j}^n\int\D v_{ij}\, v_{ij}^{-\alpha_{ij}-1} e^{-\frac{1}{4}v_{ij} x_{ij}^2}\Big)\tilde{f}(\hat{\bs{v}}). 
\end{align}

If the arbitrary function $\tilde{f}(\hat{\bs{v}}) $ is a product of powers, this expression reduces to the conformal correlator $\prod_{i<j}^n x_{ij}^{2\tilde{\alpha}_{ij}}$ where the $\tilde{\alpha}_{ij}$ satisfy $\sum_{j\neq i} \tilde{\alpha}_{ij}=-\Delta_i$.
More generally, wherever $\tilde{f}(\hat{\bs{v}})$ admits a Mellin-Barnes representation, 
we recover $\prod_{i<j}^n x_{ij}^{2\alpha_{ij}}$ times a function of position-space cross ratios as shown in  \cite{Bzowski:2020kfw}.  However, the most straightforward way to check that  \eqref{posnsimplex} solves the conformal Ward identities is to note that, when acting on a function $F=F(\{ x_{kl}^2\})$ of the squared coordinate separations, 
\begin{align}
&\sum_{i}^{n} \Big(2 x_i^\mu x_i^\nu \frac{\partial}{\partial x_i^\nu}-x_i^2 \frac{\partial}{\partial x_i^\mu}+2\Delta_i x_i^\mu\Big) F =
\sum_{i}^n 2x_i^\mu \Big(\Delta_i+\sum_{\substack{j\\j\neq i}}^n x_{ij}^2 \frac{\partial}{\partial (x_{ij}^2)} \Big) F.
\end{align}
It then follows that
\begin{align}
&\sum_{i}^{n} \Big(2 x_i^\mu x_i^\nu \frac{\partial}{\partial x_i^\nu}-x_i^2 \frac{\partial}{\partial x_i^\mu}+2\Delta_i x_i^\mu\Big) \Big(\prod_{k<l}^n\int\D v_{kl}\, v_{kl}^{-\alpha_{kl}-1} e^{-\frac{1}{4}v_{kl} x_{kl}^2}\Big)\tilde{f}(\hat{\bs{v}}) \nn\\ &\qquad
=\sum_{i}^n 2x_i^\mu \,\Big(\prod_{k<l}^n\int\D v_{kl}\, v_{kl}^{-\alpha_{kl}-1}\Big) \tilde{f}(\hat{\bs{v}}) 
\Big(\Delta_i+\sum_{\substack{j\\j\neq i}}^n v_{ij} \frac{\partial}{\partial v_{ij}} \Big)  e^{-\frac{1}{4}\sum_{k<l}^{n}v_{kl} x_{kl}^2}\nn\\
&\qquad
=\sum_{i}^n 2x_i^\mu \,\Big(\prod_{k<l}^n\int\D v_{kl}\, v_{kl}^{-\alpha_{kl}-1}\Big) \tilde{f}(\hat{\bs{v}}) 
\Big(\Delta_i+\sum_{j}^n \alpha_{ij}\Big)  e^{-\frac{1}{4}\sum_{k<l}^nv_{kl} x_{kl}^2} = 0
\end{align}
where in the last line we integrated by parts\footnote{As previously, the boundary terms vanish
provided $v_{kl}^{-\alpha_{kl}}\tilde{f}(\hat{\bs{v}})$ as $v_{kl}\rightarrow 0$. }  then used \eqref{alphaijdef}.
The middle line here accounts for the form of the total derivative we found earlier in 
\eqref{CWItotalderiv}.
Multiplying by $-i$ and Fourier transforming, the first line yields the momentum-space conformal Ward identity acting on the momentum-space simplex representation ({\it i.e.,} the left-hand side of \eqref{CWItotalderiv}), while the middle line yields
\begin{align}
& \sum_i^{n-1}2\frac{\partial}{\partial p_i^\mu}\Big(\Big(\prod_{k<l}^n \int_0^\infty \D v_{kl}\,v_{kl}^{-\alpha_{kl}-1}\Big)f(\hat{\bs{v}})\Big(\Delta_i+\sum_{\substack{j\\j\neq i}}^n v_{ij}\frac{\partial}{\partial v_{ij}}\Big) |g|^{-d/2}  e^{-\sum_{a,b}^{n-1}g^{-1}_{ab}\,\bs{p}_a\cdot\bs{p}_b}\Big)
\nn\\&=
\sum_i^{n-1}  \Big(\prod_{k<l}^n \int_0^\infty \D v_{kl}\Big)\,\sum_{\substack{j\\j\neq i}}^n \frac{\partial}{\partial v_{ij}} \Big(v_{ij} \,\Omega \,f(\hat{\bs{v}})|g|^{-d/2} \Big(\sum_a^{n-1} -4g^{-1}_{ia}p_a^\mu\Big)e^{-\sum_{a,b}^{n-1}g^{-1}_{ab}\,\bs{p}_a\cdot\bs{p}_b}\Big)
\nn\\&=
-4\sum_a^{n-1} p_a^\mu \Big(\prod_{k<l}^n \int_0^\infty \D v_{kl}\Big)\,\sum_{i\neq j}^n \frac{\partial}{\partial v_{ij}}\Big(v_{ij}\,\hat{g}^{-1}_{ia}  \Omega\, f(\hat{\bs{v}})|g|^{-d/2}e^{-\mathrm{tr}\,(g^{-1}\cdot G)}\Big)
\end{align}
where in the second line we evaluated the momentum derivative of the exponential and  pushed the factors of $\Omega=\prod_{k<l} v_{kl}^{-\alpha_{kl}-1}$, $f(\hat{\bs{v}})$ and $v_{ij}$  inside the $v_{ij}$-derivative which cancels the $\Delta_i$ term via \eqref{alphaijdef}.
In the final line, we extended the sum over $i$ to run up to $n$ by replacing $g^{-1}_{ia}$ with $\hat{g}^{-1}_{ia}$ and combined it with the sum over $j$.  Up to a relabelling of indices, this final line is now the total derivative appearing on the right-hand side of \eqref{CWItotalderiv}.

The manipulations above illustrate a  general theme: given
the simplicity of the position-space simplex representation \eqref{posnsimplex}, it is often  profitable to work with the position-space equivalents of differential operators in order to evaluate their action in terms of the $v_{ij}$ variables.  Both sides can then be Fourier transformed back to momentum space in order to deduce the action of the corresponding momentum-space operator on the momentum-space simplex in terms of the $v_{ij}$ variables.   In many cases this is more straightforward than working in momentum space throughout.

\subsection[\texorpdfstring{Action of $\mathcal{W}_{12}^{-+}$}{Action of W(-+)}]{\texorpdfstring{Action of \boldmath{$\mathcal{W}_{12}^{-+}$}}{Action of W(-+)}}
\label{sec:Wmp}

As a further illustration of this approach, let us evaluate the action of the shift operator $\mathcal{W}_{12}^{-+}$ defined in \eqref{Wops}.
After expanding out the derivative, this operator can easily be Fourier transformed to position space where it reads
\[
\mathcal{W}_{12}^{-+}=\frac{1}{2}x_{12}^2 \frac{\partial}{\partial x_2^\mu}\frac{\partial}{\partial x_{2\mu}}+2(\beta_2+1)\Big(\beta_2+\frac{d}{2}-x_{12}^\mu \frac{\partial}{\partial x_2^\mu}\Big).
\]
Acting on a function $F=F(\{ x_{kl}^2\})$ of the squared coordinate separations, we find via the chain rule
\begin{align}
\mathcal{W}_{12}^{-+} F &= 
\sum_{\substack{i,j\\i,j\neq 2}}^n x_{12}^2 (x_{2i}^2+x_{2j}^2-x_{ij}^2)\frac{\partial^2F}{\partial (x_{2i}^2)\partial (x_{2j}^2)}\nn\\&\quad
+\sum_{i\neq 2}^n\Big(2(\beta_2+1)(x_{12}^2-x_{1i}^2+x_{2i}^2)+dx_{12}^2\Big)\frac{\partial F}{\partial (x_{2i}^2)}
+2(\beta_2+1)\Big(\beta_2+\frac{d}{2}\Big)F
\end{align}
Acting on the Schwinger exponent appearing in the position-space simplex representation \eqref{posnsimplex}, this can be translated into $v_{ij}$-derivatives  as
\begin{align}
&\mathcal{W}_{12}^{-+} \Big(\prod_{k<l}^n\int\D v_{kl}\, v_{kl}^{-\alpha_{kl}-1} e^{-\frac{1}{4}v_{kl} x_{kl}^2}\Big)\tilde{f}(\hat{\bs{v}}) \nn\\
&\quad = 2\,\Big(\prod_{k<l}^n\int\D v_{kl}\, v_{kl}^{-\alpha_{kl}-1} \Big)\tilde{f}(\hat{\bs{v}}) \Big[\,(\beta_2+1)\Big(\beta_2+\frac{d}{2}\Big)+
\Big(2\beta_2+1+\frac{d}{2}+\theta_{v_{12}}\Big)\theta_{v_{12}}
\nn\\&\qquad\qquad\qquad
+\sum_{i=3}^n  v_{2i}\Big((\beta_2+1+\theta_{v_{12}})(\partial_{v_{12}}+\partial_{v_{2i}}-\partial_{v_{1i}})+\Big(\frac{d}{2}+\theta_{v_{2i}}\Big)\partial_{v_{12}}\Big)\nn\\&\qquad\qquad\qquad
+\sum_{3\le i<j}^n v_{2i}v_{2j}(\partial_{v_{2i}}+\partial_{v_{2j}}-\partial_{v_{ij}})\partial_{v_{12}}\Big] e^{-\frac{1}{4}\sum_{k<l}^nv_{kl} x_{kl}^2}
\end{align}
where $\partial_{v_{ij}}=\partial/\partial v_{ij}$ and $\theta_{v_{ij}} = v_{ij}\partial_{v_{ij}}$. 
Integrating by parts, we find
\begin{align}
&\mathcal{W}_{12}^{-+} \Big(\prod_{k<l}^n\int\D v_{kl}\, v_{kl}^{-\alpha_{kl}-1} e^{-\frac{1}{4}v_{kl} x_{kl}^2}\Big)\tilde{f}(\hat{\bs{v}}) \nn\\&
\quad = 2\,\Big(\prod_{k<l}^n\int\D v_{kl}\,e^{-\frac{1}{4}v_{kl} x_{kl}^2}\Big)
\Big[(\theta_{v_{12}}-\beta_2)\Big(\theta_{v_{12}}-\beta_2-\frac{d}{2}-1+n\Big)
\nn\\&\qquad\qquad
+\sum_{i=3}^n v_{2i}\Big((\theta_{v_{12}}-\beta_2)(\partial_{v_{12}}+\partial_{v_{2i}}-\partial_{v_{1i}})+\Big(n-\frac{d}{2}+\theta_{v_{2i}}\Big)\partial_{v_{12}}\Big)\nn\\&\qquad
\qquad+
\sum_{3\le i<j}^n v_{2i}v_{2j}(\partial_{v_{2i}}+\partial_{v_{2j}}-\partial_{v_{ij}})\partial_{v_{12}}\Big] \Omega \tilde{f}(\hat{\bs{v}}).
\label{step0}
\end{align}
We now rewrite the first part of the last line as 
\begin{align}
&\Big[\sum_{3\le i<j}^n v_{2i}v_{2j}(\partial_{v_{2i}}+\partial_{v_{2j}})\partial_{v_{12}}\Big]\Omega \tilde{f}(\hat{\bs{v}})
=\Big[\sum_{i=3}^n \Big(\sum_{\substack{j=3\\j\neq i}}^n v_{2j}\Big)\theta_{v_{2i}}\partial_{v_{12}}
\Big]\Omega \tilde{f}(\hat{\bs{v}})\nn\\&\qquad 
=\Big[
\Big(\sum_{j=3}^n v_{2j}\Big)\partial_{v_{12}}\Big(-\theta_{v_{12}}+\sum_{i\neq 2}^n \theta_{v_{2i}}\Big)-\sum_{i=3}^n v_{2i}\theta_{v_{2i}}\partial_{v_{12}}\Big]\Omega \tilde{f}(\hat{\bs{v}})\nn\\&\qquad 
=\Big[-
\sum_{i=3}^n v_{2i}\Big((\theta_{v_{12}} +1)-\beta_2-\frac{d}{2}+(n-1)+
\theta_{v_{2i}}\Big)\partial_{v_{12}}\Big]\Omega \tilde{f}(\hat{\bs{v}})
\label{step1}
\end{align}
where in the final step we rewrote $\partial_{v_{12}}\theta_{v_{12}}=(\theta_{v_{12}}+1)\partial_{v_{12}}$ and
used $\sum_{i\neq 2}^n\theta_{v_{2i}}\tilde{f}(\hat{\bs{v}})=0$, as follows from \eqref{sumofEuleronf}, along with \eqref{alphaijdef} with $\Delta_2=\beta_2+d/2$ to replace
\[\label{step2}
\Big(\sum_{i\neq 2}^n\theta_{v_{2i}}\Big)\Omega \tilde{f}(\hat{\bs{v}})=(\beta_2+d/2-(n-1))\Omega  \tilde{f}(\hat{\bs{v}}).
\]
Substituting \eqref{step1} into \eqref{step0} and making further use of \eqref{step2}, we find the result
\begin{align}
&\mathcal{W}_{12}^{-+} \Big(\prod_{k<l}^n\int\D v_{kl}\, v_{kl}^{-\alpha_{kl}-1} e^{-\frac{1}{4}v_{kl} x_{kl}^2}\Big)\tilde{f}(\hat{\bs{v}}) \nn\\&
\quad = -2\,\Big(\prod_{k<l}^n\int\D v_{kl}\,e^{-\frac{1}{4}v_{kl} x_{kl}^2}\Big)
\Big[
(\theta_{v_{12}}-\beta_2)\sum_{i=3}^n v_{2i}\partial_{v_{1i}}
+
\sum_{3\le i<j}^n v_{2i}v_{2j}\partial_{v_{ij}}\partial_{v_{12}}\Big] \Omega \tilde{f}(\hat{\bs{v}}).
\end{align}
Equivalently, acting on the position-space simplex with $\mathcal{W}_{12}^{-+}$ corresponds to acting on the arbitrary function $\tilde{f}(\hat{\bs{v}})$ with the operator
\[\label{tildeWdef}
\tilde{\mathcal{W}}_{12}^{-+} =-2 \Omega^{-1}\Big[
(\theta_{v_{12}}-\beta_2)\sum_{i=3}^n v_{2i}\partial_{v_{1i}}
+
\sum_{3\le i<j}^n v_{2i}v_{2j}\partial_{v_{ij}}\partial_{v_{12}}\Big] \Omega. 
\]
The same remains true when we Fourier transform back to momentum space, giving
\begin{align}
&\mathcal{W}_{12}^{-+}\Big(\prod_{k<l}^n\int_0^\infty\D v_{kl}\, v_{kl}^{-\alpha_{kl}-1}\Big)|g|^{-d/2}e^{-\mathrm{tr}(g^{-1}\cdot G)}f(\hat{\bs{v}})\nn\\&\qquad 
=\Big(\prod_{k<l}^n\int_0^\infty\D v_{kl}\, v_{kl}^{-\alpha_{kl}-1}\Big)|g|^{-d/2}e^{-\mathrm{tr}(g^{-1}\cdot G)}(\tilde{\mathcal{W}}_{12}^{-+}f(\hat{\bs{v}})).
\end{align}

Finally, it remains to check that the action of $\tilde{\mathcal{W}}_{12}^{-+}$ on the arbitrary function produces the required shift in dimensions $\Delta_1\rightarrow \Delta_1-1$ and $\Delta_2\rightarrow \Delta_2+1$.  
Since 
\[
\partial_{v_{ij}}\Omega=-(\alpha_{ij}+1)\frac{\Omega}{v_{ij}},\qquad \partial_{v_{ij}}f(\hat{\bs{v}})=\frac{ h(\hat{\bs{v}})
}{v_{ij}},
\] 
where $h(\hat{\bs{v}})$ is also function of the cross ratios, we see that 
\begin{align}\label{tildeWaction}
&\tilde{\mathcal{W}}_{12}^{-+} f(\hat{\bs{v}}) = 
\sum_{i=3}^n \frac{v_{2i}}{v_{1i}} h_i(\hat{\bs{v}})
+
\sum_{3\le i<j}^n \frac{v_{2i}v_{2j}}{v_{ij}v_{12}}h_{ij}(\hat{\bs{v}})
\end{align}
where $h_i(\hat{\bs{v}})$ and $h_{ij}(\hat{\bs{v}})$ are specific functions of the cross ratios.   Each term in the first sum then corresponds to a simplex integral with the shifts 
\[\label{shiftpattern1}
\alpha_{2i}\rightarrow \alpha_{2i}-1,\qquad
\alpha_{1i}\rightarrow \alpha_{1i}+1,
\]
while each term in the second sum  corresponds to a simplex integral with the shifts
\[\label{shiftpattern2}
\alpha_{2i}\rightarrow \alpha_{2i}-1, \qquad \alpha_{2j}\rightarrow \alpha_{2j}-1,\qquad
\alpha_{ij}\rightarrow \alpha_{ij}+1,\qquad
\alpha_{12}\rightarrow \alpha_{12}+1.
\]
From \eqref{alphaijdef}, both \eqref{shiftpattern1} and \eqref{shiftpattern2}  correspond to shifting $\Delta_1\rightarrow \Delta_1-1$ and $\Delta_2\rightarrow \Delta_2+1$ leaving all other operator dimensions fixed. 
The action of $\mathcal{W}_{12}^{-+}$ on the simplex thus produces an appropriately shifted simplex integral, whose function of cross ratios is obtained through the action of the operator \eqref{tildeWdef}.

\section{Discussion}
\label{sec:disc}

Our analysis has furnished
useful
parametric representations for the general momentum-space conformal $n$-point function.  
Starting from the generalised simplex Feynman integral of \cite{Bzowski:2019kwd, Bzowski:2020kfw}, 
we showed how all graph polynomials can be obtained from the corresponding Laplacian matrix, or the Gram matrix to which it reduces once momentum conservation has been enforced. 
With the graph polynomials to hand, all the usual scalar parametrisations of Feynman integrals can  be adapted to represent the simplex solution.   Only $n(n-1)/2$ integrals over Schwinger parameters remain to be performed --  one for each leg of the simplex -- in contrast to the $(n-1)(n-2)d/2$ scalar integrals we started with.

Building on  the analogy between Feynman graph polynomials and those  of electrical circuits, we then formulated a second class of parametric representations.  For these, the integration variables represent the {\it effective} resistances between vertices of the simplex, 
rather than   
the conductivities 
({\it i.e.,} the inverse Schwinger parameters) used previously.  
This change of variables immediately diagonalises the Schwinger exponential, expressing the $n$-point function as a standard Laplace transform of a product of polynomials raised to generalised powers.  These polynomials correspond to the determinant and  first minors of the Cayley-Menger matrix for the simplex, which 
plays an analogous role to the Gram matrix for this second class of parametrisations.  From the form of these polynomials, new weight-shifting operators can  immediately be constructed to raise the power of these polynomials, with further shift operators  following by shadow conjugation.   Besides shifting the scaling dimensions of external operators, these new weight-shifting operators  raise the spacetime dimension by two.
They therefore generalise the $3$-point shift operators of \cite{Bzowski:2013sza, Bzowski:2015yxv} to $n$-points, and 
 constitute a distinct class of operators to those 
 identified in  \cite{Karateev:2017jgd}.

Our results suggest several interesting directions for further pursuit:
\begin{itemize}
\item Given  we now have  weight-shifting operators that both preserve and raise the spacetime dimension, is it also possible to construct operators that {\it lower} the spacetime dimension?  
One approach we have explored, explained in appendix \ref{app:Bernstein},  is to find   so-called {\it Bernstein-Sato} operators which act to lower the powers to which the  various polynomials of interest are raised.  In this case, the relevant polynomials are
the Cayley-Menger determinant and its minors appearing in the parametrisation \eqref{Cay2}.  
We found, for example,  that  replacing $v_{ij}\rightarrow\partial_{s_{ij}}$ in the Kirchhoff polynomial $\mathcal{U}=|g|$ yields an operator 
\[
\mathcal{B}_{|m|} = (|g|)\Big|_{v_{ij}\rightarrow \partial_{s_{ij}}}
\]
which lowers by one the power to which the Cayley-Menger determinant is raised: 
\[\label{CMBernstein}
\mathcal{B}_{|m|}\, |m|^a = b_{|m|}(a) |m|^{a-1}, \qquad b_{|m|}(a) =- \prod_{k=1}^{n-1}(1-k-2a).
\] 
For the simplex representation \eqref{Cay2}, $a$ is the parameter $\alpha$ given in \eqref{alphadef} and so lowering $\alpha$ by one corresponds to sending $d\rightarrow d-2$ if all the operator dimensions are kept fixed.   In principle, one would then integrate by parts to obtain an operator acting solely on the Schwinger exponential, which, due to its diagonal structure, could be 
translated into a differential operator in the external momenta.
In practice, however, this approach 
 is complicated by the presence of all the 
remaining powers of Cayley-Menger minors present in \eqref{Cay2}.

\item In sections \ref{sec:CWI} and \ref{sec:posnspacecwis}, we saw how the action of the special conformal Ward identity on the simplex reduces  to a total derivative.   This followed directly from the scalar parametric representation, without any recourse to the recursive arguments developed in \cite{Bzowski:2019kwd, Bzowski:2020kfw}.
Nevertheless, these arguments, and the recursion relation between  $n$- and $(n+1)$-point simplices on which they are based, are  of  considerable
interest in their own right and could be reformulated in the scalar-parametric language used here.
The deletion/contraction relations of graph polynomials (see, {\it e.g.,} \cite{Weinzierl:2022eaz}) and Kron reduction, corresponding to taking the Schur complement of a subset of vertices in the simplex Laplacian (see {\it e.g.,} \cite{dorfler2012kron}), may also yield relevant identities.

\item Starting from the general simplex solution, the arbitrary function of momentum-space cross ratios can be restricted by imposing additional conditions of interest: for example, dual conformal invariance \cite{Bzowski:2015pba, Coriano:2019sth, Loebbert:2020hxk, Rigatos:2022eos}, or the Casimir equation for conformal blocks.  For such investigations, the connection with position-space developed in section \ref{sec:posnspace}  provides a very simple link between the action of a given differential operator in the external momenta or coordinates, and its corresponding action on the arbitrary function of the simplex representation.

\item For holographic $n$-point functions, bulk scalar Witten diagrams have the interesting property that their form is invariant under the action of a shadow transform on any of the external legs.  In momentum space, shadow transforming the operator $\O_{i}$ corresponds to multiplying the correlator by $p_i^{-2\beta_i}$, where $\beta_i=\Delta_i-d/2$, which  has the effect of replacing $\beta_i\rightarrow - \beta_i$ in the   bulk-boundary propagator 
$z^{d/2} p_i^{\beta_i} K_{\beta_i}(p_iz)$.  
It would be interesting to understand the restriction this condition places on the function of cross-ratios appearing in the  simplex representation.

\item Finally, the  parametric  representations we have developed may provide a useful starting point for  the construction of general spinning $n$-point correlators via the action of spin-raising operators \cite{Karateev:2017jgd, Arkani-Hamed:2018kmz, Baumann:2019oyu}, and for bootstrapping cosmological correlators in de Sitter spacetime.   

\end{itemize}

\acknowledgments

FC thanks the School of Maths, Statistics \& Physics at Newcastle University for support.  PM is supported by  the UKRI 
through the Ernest Rutherford Fellowship ST/P004326/2.

\appendix

\section{Derivation of graph polynomials}
\label{incidence_app}

In this appendix, we show that Schwinger representation of the simplex integral \eqref{simplex} is given by \eqref{Schwrep} with   graph polynomials given in \eqref{Kirchhoff1}.  Our discussion builds on that in  \cite{Itzykson:1980rh}.
Labelling the vertices of the simplex by $i=1,\ldots,n$, and the (directed) legs by $a=1,\ldots,N$ where $N=n(n-1)/2$, we introduce the incidence matrix 
\[
\ep_i^a = \begin{cases} +1\qquad &\mathrm{if\, leg}\, a\, \mathrm{is\, ingoing\, to\, vertex}\, i\\
-1\qquad& \mathrm{if\, leg}\, a\, \mathrm{is\, outgoing\, to\, vertex}\, i\\
0\qquad &\mathrm{otherwise}
\end{cases}
\]
where for clarity we will write the vertex index downstairs and the leg index upstairs. 
Thus, for example, if we choose $a=\{(12),(13),(14),(23),(24),(34)\}$ as the legs of the 4-point function, where the leg $(i,j)$ runs from vertex $i$ to vertex $j$, the incidence matrix is 
\[
\ep = \left(\begin{matrix}
-1 & -1 & -1 &0 &0 &0\\
1&0&0&-1&-1&0\\
0&1&0&1&0&-1\\
0&0&1&0&1&1
\end{matrix}\right).
\]
Momentum conservation at vertex $k$ of the simplex can now be re-expressed as 
\[
0=\bs{p}_k + \sum_{l\neq k}^n \bs{q}_{lk}=\bs{p}_k + \sum_{a}^N \ep_k^a\,\bs{q}_{a}
\]
where $\bs{q}_a$ is the internal momentum flowing along the directed leg $a$.  As always,  all sums are assumed to begin at one unless otherwise specified. 
The Laplacian matrix $\tilde{g}_{ij}$ defined in \eqref{tildeg} can now be written 
\[\label{Laplep}
\tilde{g}_{ij} = \sum_a^N v_a \ep_i^a \ep_j^a,
\]
which follows by noting that for $i\neq j$ only the leg for which $a$ runs between vertices $i$ and $j$ contributes giving $-v_{ij}$, while for $i=j$ all legs running into this vertex contribute giving $\sum_{k\neq i}v_{ik}$ as required. 

Turning to the simplex integral \eqref{simplex}, we first rewrite the delta functions of momentum conservation in Fourier form 
\[
\prod_{k}^n \,(2\pi)^d\delta\Big(\bs{p}_k + \sum_{l\neq k}^n \bs{q}_{lk}\Big) = \prod_k^n \int \mathrm{d}^d\bs{y}_k \, \exp\Big(-i\bs{y}_k\cdot\big(\bs{p}_k + \sum_{a}^N \ep_k^a\,\bs{q}_{a}\big)\Big).
\]
Next, exponentiating all propagators of internal momenta (labelled by their leg indices) using the Schwinger representation \eqref{Schwingerintegral}, we find\footnote{Note the argument of the arbitrary function $f$ changes from the momentum cross ratios $\hat{\bs{q}}$ in \eqref{conf_ratio_q}  to the Schwinger parameter cross ratios $\hat{\bs{v}}$ in \eqref{vcrossratio}.   This can be seen by temporarily representing the arbitrary function in Mellin-Barnes form ({\it i.e.,} (4.18) of \cite{Bzowski:2020kfw}) allowing all $q_{ij}$, including those from the cross ratios, to be exponentiated via the Schwinger parametrisation \eqref{Schwingerintegral}.  Performing the Mellin-Barnes integration then generates $f(\hat{\bs{v}})$, since the Schwinger parametrisation replaces powers of $q_{ij}$  by powers of $v_{ij}$.}
\begin{align}
&\< \O_1(\bs{p}_1) \ldots \O_n(\bs{p}_n) \>
=\Big(\prod_a^N \frac{1}{\Gamma(\alpha_a+d/2)}\int_0^\infty \D v_a\,
v_a^{-d/2-\alpha_a-1}\Big)f(\hat{\bs{v}}) \nn\\&\qquad\times
\Big(\prod_{k}^n \int\D^d\bs{y}_k \, \exp(-i\bs{y}_k\cdot\bs{p}_k)
\Big)\int\frac{\D^d\bs{q}_a}{(2\pi)^d} \exp\Big(-\sum_a^N\Big(\frac{q_a^2}{v_a}+i\sum_l^n\ep_l^a\bs{y}_l\cdot \bs{q}_a\Big)\Big).
\end{align}
Evaluating the $\bs{q}_a$ integrals by completing the square and using \eqref{Laplep} now gives
\begin{align}
\< \O_1(\bs{p}_1) \ldots \O_n(\bs{p}_n) \>&
=\Big(\prod_a^N \frac{\pi^{d/2}}{\Gamma(\alpha_a+d/2)}\int_0^\infty \D v_a\,
v_a^{-\alpha_a-1}\Big)f(\hat{\bs{v}}) \nn\\&\qquad\times
\Big(\prod_{k}^n \int\D^d\bs{y}_k \Big)\, \exp\Big(-i\sum_k^n \bs{y}_k\cdot\bs{p}_k - \frac{1}{4}\sum_{k,l}^n \tilde{g}_{kl}\,\bs{y}_k\cdot\bs{y}_l\Big).
\end{align}
Since the Laplacian matrix has no inverse, to compute the $\bs{y}_k$ integrals we must first shift
\begin{align}
 \bs{y}_n = \bs{z}_n, \qquad \bs{y}_{k}=\bs{z}_k +\bs{z}_n, \qquad k=1,\ldots n-1.
\end{align}
This transformation has unit Jacobian, but moreover greatly simplifies the exponent.   Since all row and column sums of the Laplacian matrix vanish, 
\[
\sum_{l}^{n-1} \tilde{g}_{kl} = -\tilde{g}_{kn},\qquad\qquad  \sum_k^{n-1}\tilde{g}_{kn} = -\tilde{g}_{nn},
\]
and using these identities we then find
\begin{align}
&-i\sum_k^n \bs{y}_k\cdot\bs{p}_k - \frac{1}{4}\sum_{k,l}^n \tilde{g}_{kl}\,\bs{y}_k\cdot\bs{y}_l \nn\\&\qquad 
=-i\bs{z}_n\cdot\bs{p}_n -i\sum_k^{n-1} (\bs{z}_k+\bs{z}_n)\cdot\bs{p}_k \nn\\&\qquad\quad
-\frac{1}{4}\tilde{g}_{nn}z_n^2-\frac{1}{2}\sum_k^{n-1}\tilde{g}_{kn}(\bs{z}_k+\bs{z}_n)\cdot\bs{z}_n - \frac{1}{4}\sum_{k,l}^{n-1} \tilde{g}_{kl}\,(\bs{z}_k+\bs{z}_n)\cdot(\bs{z}_l+\bs{z}_n)
\nn\\&\qquad=
-i\bs{z}_n\cdot\Big(\sum_k^n\bs{p}_k\Big) -i\sum_k^{n-1}\bs{z}_k\cdot\bs{p}_k -\frac{1}{4}\sum_{k,l}^{n-1}g_{kl}\,\bs{z}_k\cdot\bs{z}_l.
\end{align}
In the final line here, all the $\bs{z}_k\cdot\bs{z}_n$ and $z_n^2$ terms cancel while the Laplacian matrix $\tilde{g}_{kl}$ reduces to $g_{kl}$ for $k,l=1,\ldots ,n$.   The $\bs{z}_n$ integral now gives the overall delta function of momentum conservation which we  strip off to obtain the reduced correlator \eqref{redcorr}.  The remaining $\bs{z}_k$ integrals can be evaluated by completing the square, given that the inverse $g_{kl}^{-1}$ exists. 
This yields our desired result,
\begin{align}
&\lla \O_1(\bs{p}_1) \ldots \O_n(\bs{p}_n) \rra = 
\mathcal{C}\prod_a^N \int_0^\infty\D v_a\, v_a^{-\alpha_a-1} f(\hat{\bs{v}})\, |g|^{-d/2} \exp\Big(-\sum_{k,l}^{n-1}g_{kl}^{-1}\,\bs{p}_k\cdot\bs{p}_l\Big)
\end{align}
where the constant
\[
\mathcal{C} =(4\pi)^{(n-1)d/2} \prod_a^N \frac{\pi^{d/2}}{\Gamma(\alpha_a+d/2)}
\]
can simply be re-absorbed into the arbitrary function $f(\hat{\bs{v}})$.
Rewriting the product of legs $a$ as a product over vertices $i<j$ 
and replacing $\bs{p}_k\cdot\bs{p}_l$ with the Gram matrix $G_{kl}$, we recover precisely \eqref{Schwrep} with graph polynomials \eqref{Kirchhoff1}.

\section{Jacobian matrix} 

In this appendix, we compute the Jacobian matrix for the change of variables from $v_{ij}$ to $s_{ij}$.  In section \ref{Jac_app} we evaluate the Jacobian determinant, then in section \ref{sec:Jacobimatrixels}
we give expressions for its matrix elements enabling conversion between partial derivatives.

\subsection{Jacobian determinant}
\label{Jac_app}

Our first goal is to derive the relation \eqref{Jac0} for  the Jacobian determinant, namely 
\begin{align}\label{Jac1}
\left|\frac{\partial s}{\partial v}\right| = \left|\frac{\partial^2 \ln |g|}{\partial v\,\partial v}\right| \propto |g|^{-n},
\end{align}
where the constant of proportionality is not required since it can be re-absorbed into the arbitrary function $f(\hat{\bs{v}})$.
For small values of $n$ this result can be verified by direct calculation, and the exponent is simply fixed  by power counting, but our aim  is nevertheless to prove this relation for general $n$.

We start by noting
\begin{align}\label{matrixprod}
\frac{\partial^2 \ln |g|}{\partial v_{ij}\partial v_{kl}} = \frac{\partial g_{pq}}{\partial v_{ij}}\frac{\partial^2 \ln |g|}{\partial g_{pq}\partial g_{rs}} \frac{\partial g_{rs}}{\partial v_{kl}}
\end{align}
can be re-expressed as a product of three square matrices of dimension $n(n-1)/2$.  
Each of the index pairs $(p,q)$ and $(r,s)$ is replaced by a single index running over  the $n(n-1)/2$ independent entries of the  $(n-1)\times(n-1)$ symmetric matrix $g$, while $(i,j)$ and $(k,l)$ are each replaced by a single index running over the $n(n-1)/2$ edges of the simplex.  Noting the elements of $g$ are linear in the $v$, the matrix determinant $|\partial g/\partial v|$ evaluates to a nonzero constant.  On taking the determinant of \eqref{matrixprod}, we find 
\[
 \left|\frac{\partial^2 \ln |g|}{\partial v\,\partial v}\right| \propto  \left|\frac{\partial^2 \ln |g|}{\partial g \,\partial g }\right|
\]
hence it suffices to show that
\[\label{Hess0}
 \left|\frac{\partial^2 \ln |g|}{\partial g \,\partial g }\right| \propto |g|^{-n}.
\]
This relation in fact holds for any invertible symmetric square matrix  $g$ of dimension $n-1$.

To see this, from Jacobi's relation we have
\[\label{Hess1}
\frac{\partial^2 \ln |g|}{\partial g_{pq}\partial g_{rs}}
=\frac{\partial}{\partial g_{rs}}\Big(\frac{1}{|g|}(\mathrm{adj}\, g)_{pq}\Big)  = \frac{\partial (g^{-1})_{pq}}{\partial g_{rs}}=-2(g^{-1})_{p(r}(g^{-1})_{s)q}.
\]
Diagonalising $g$ via an orthogonal matrix $O$, 
\[
\Lambda = O g O^{-1},
\] 
since
$
g^{-1} = O^{-1}\Lambda^{-1}O
$
the chain rule gives 
\[
\frac{\partial g^{-1}}{\partial{g}} = \frac{\partial (O^{-1}\Lambda^{-1}O)}{\partial \Lambda^{-1}}\frac{\partial\Lambda^{-1}}{\partial \Lambda}\frac{\partial(OgO^{-1})}{\partial g}
\]
where the last factor is just $\partial \Lambda/\partial g$.
Regarding this as a matrix product, the first and last matrices depend only on $O$ and are inverses of each other.   On taking the determinant of the right-hand side, their contributions therefore cancel giving
\[
\left|\frac{\partial g^{-1}}{\partial{g}} \right| = \left|\frac{\partial\Lambda^{-1}}{\partial \Lambda}\right|.
\]
We thus only need to evaluate the latter determinant for the diagonal matrix $\Lambda$.

From \eqref{Hess1}, the Hessian 
is nonzero only when the index pairs are equal $(p,q)=(r,s)$, and  is thus diagonal when regarded as a square matrix of dimension $n(n-1)/2$:
\begin{align}
\frac{\partial^2\ln |\Lambda|}{\partial \Lambda_{pq}\partial \Lambda_{rs}} = \frac{\partial (\Lambda^{-1})_{pq}}{\partial \Lambda_{rs}}=\begin{cases} -\Lambda_{pp}^{-1}\Lambda_{qq}^{-1} \qquad &\mathrm{if}\quad (p,q) = (r,s) \\
0 \qquad& \mathrm{otherwise}\end{cases}
\end{align}
The determinant is now
\begin{align}
\left|\frac{\partial^2\ln |\Lambda|}{\partial \Lambda_{pq}\Lambda_{rs}}\right|\,
\propto \,\prod_{p=1}^{n-1} (\Lambda_{pp})^{-n} = |\Lambda|^{-n} = |g|^{-n}
\end{align}
since each eigenvalue $\Lambda_{pp}$ appears a total of $n$ times along the diagonal: for example, $\Lambda_{11}$ appears  quadratically in  the position $(1,1)$ and then linearly in each of the $(n-2)$ entries indexed by $(1,q)$ for $q=2,\ldots n-1$. 
We have thus established \eqref{Hess0}, and hence \eqref{Jac1}.

\subsection{Matrix elements}
\label{sec:Jacobimatrixels}

We now compute the elements of the Jacobian matrix required to establish the relation
\[\label{dvinds}
\partial_{v_{kl}} = -\frac{1}{4}\sum_{i<j}^n(s_{ki}-s_{li}-s_{kj}+s_{lj})^2\partial_{s_{ij}}
\]
which we used in \eqref{Wmmid}.
Starting with \eqref{srel1},
\begin{align}
\frac{\partial s_{ij}}{\partial v_{kl}} = \frac{\partial}{\partial v_{kl}}\Big((g^{-1})_{ab}\frac{\partial g_{ab}}{\partial v_{ij}}\Big) =\frac{\partial (g^{-1})_{ab}}{\partial v_{kl}}\frac{\partial g_{ab}}{\partial v_{ij}}
\end{align}
where since $g_{ab}$ is linear in the $v_{ij}$ its second derivative vanishes.  Using
\[
\frac{\partial (g^{-1})_{ab} }{\partial v_{kl}} = - (g^{-1})_{ae}(g^{-1})_{bf}\frac{\partial g_{ef} }{\partial v_{kl}} 
\]
then gives 
\begin{align}
\frac{\partial s_{ij}}{\partial v_{kl}} &= -\mathrm{tr}\Big(g^{-1}\cdot \frac{\partial g}{\partial v_{ij}}\cdot g^{-1}\cdot\frac{\partial g}{\partial v_{jk}}\Big) =-\sum_{a,b,e,f}^{n-1}(g^{-1})_{ae}\frac{\partial g_{ef}}{\partial v_{kl}}(g^{-1})_{fb}\frac{\partial g_{ba}}{\partial v_{ij}}. \label{dsdv1}
\end{align}
For $i,j,k,l \neq n$, we can evaluate this as 
\begin{align}
&\frac{\partial s_{ij}}{\partial v_{kl}} 
= -\sum_{a,b,e,f}^{n-1}(g^{-1})_{ae}(-2\delta_{k(e}\delta_{f)l}
+\delta_{ek}\delta_{fk}+\delta_{el}\delta_{fl}))
(g^{-1})_{fb}(-2\delta_{i(a}\delta_{b)j}+\delta_{ai}\delta_{bi}+\delta_{aj}\delta_{bj})\nn\\
&=-2(g^{-1})_{ik}(g^{-1})_{jl}-2(g^{-1})_{il}(g^{-1})_{jk}+2(g^{-1})_{ik}(g^{-1})_{jk}+2(g^{-1})_{il}(g^{-1})_{jl}\nn\\&
\quad +2 (g^{-1})_{ik}(g^{-1})_{il}+2(g^{-1})_{jk}(g^{-1})_{jl}
-((g^{-1})_{ik})^2-((g^{-1})_{il})^2-((g^{-1})_{jk})^2-((g^{-1})_{jl})^2\nn\\[1ex]
&= -\big((g^{-1})_{ik}-(g^{-1})_{il}-(g^{-1})_{jk}+(g^{-1})_{jl})\big)^2\nn\\&
=-\frac{1}{4}\big(s_{ik}-s_{il}-s_{jk}+s_{jl}\big)^2\label{srel2}
\end{align}
where we used the symmetry of the inverse matrix $g^{-1}_{ij}$, and in the last line we used \eqref{ginvs}.
For $j=n$ but $i,k,l\neq n$, 
\begin{align}
\frac{\partial s_{in}}{\partial v_{kl}} 
&= -\sum_{a,b,e,f}^{n-1}(g^{-1})_{ae}(-2\delta_{k(e}\delta_{f)l}
+\delta_{ek}\delta_{fk}+\delta_{el}\delta_{fl}))
(g^{-1})_{fb}(\delta_{ai}\delta_{bi})\nn\\
&
=-\big((g^{-1})_{ik}- (g^{-1})_{il}\big)^2
=-\frac{1}{4}\big(s_{ik}-s_{il}-s_{kn}+s_{ln}\big)^2
\end{align}
which is equivalent to \eqref{srel2} setting $j=n$.
The same also holds for $l=n$ but $i,j,k\neq n$ due to the symmetry of \eqref{dsdv1}.  Finally
\begin{align}
\frac{\partial s_{in}}{\partial v_{kn}} 
&= -\sum_{a,b,e,f}^{n-1}(g^{-1})_{ae}(\delta_{ek}\delta_{fk}))
(g^{-1})_{fb}(\delta_{ai}\delta_{bi})\nn\\&
=-((g^{-1})_{ik})^2=-\frac{1}{4}(s_{ik}-s_{in}-s_{kn})^2,
\end{align}
also equivalent to \eqref{srel2} since $s_{nn}=0$.
Thus \eqref{srel2} in fact holds for all values of the indices and we obtain \eqref{dvinds}.

For completeness, we can also calculate the inverse Jacobian by similar means:
\begin{align}
\frac{\partial v_{ij}}{\partial s_{kl}} &= \frac{\partial}{\partial s_{kl}}\Big( (m^{-1})_{ab}\frac{\partial m_{ab}}{\partial s_{ij}}\Big) = \frac{\partial (m^{-1})_{ab}}{\partial s_{kl}}\frac{\partial m_{ab}}{\partial s_{ij}}=-(m^{-1})_{ae}\frac{\partial m_{ef}}{\partial s_{kl}}(m^{-1})_{fb}\frac{\partial m_{ba}}{\partial s_{ij}}\nn\\
&=-(m^{-1})_{ae}(\delta_{ek}\delta_{fl}+\delta_{el}\delta_{fk})(m^{-1})_{fb}(\delta_{bi}\delta_{aj}+\delta_{bj}\delta_{ai})\nn\\&
=-2\Big((m^{-1})_{ik}(m^{-1})_{jl}+(m^{-1})_{li}(m^{-1})_{jk}\Big).
\label{dvds1}
\end{align}
Apart from the final $(n+1)^{\mathrm{th}}$ row and column, the inverse Cayley-Menger matrix is  minus one half the Laplacian matrix $\tilde{g}_{ij}$ as we showed in \eqref{minvisLa1} and \eqref{minvisLa2}.  This gives
\[
\frac{\partial v_{ij}}{\partial s_{kl}} =-\frac{1}{2}\Big(\tilde{g}_{ik}\tilde{g}_{jl}+\tilde{g}_{il}\tilde{g}_{jk}\Big),\label{dvds}
\]
where $\tilde{g}_{ij}=-v_{ij}$ for $i\neq j$ and $\tilde{g}_{ii}=\sum_{a=1}^n v_{ia}$. 
For $i,j,k,l$ all different, we therefore have 
\[
\frac{\partial v_{ij}}{\partial s_{kl}} =-\frac{1}{2}\Big(v_{ik}v_{jl}+v_{il}v_{jk}\Big), \qquad i\neq j\neq k\neq l
\]
while if $j=l$, 
\[
\frac{\partial v_{ij}}{\partial s_{kj}} =-\frac{1}{2}\Big(v_{ij}v_{jk}-v_{ik}\Big(\sum_{a=1}^n v_{ja}\Big)\Big)
\]
and if $i=k$ and $j=l$,
\[
\frac{\partial v_{ij}}{\partial s_{ij}} =-\frac{1}{2}\Big(v_{ij}^2+\Big(\sum_{a=1}^n v_{ia}\Big)\Big(\sum_{b=1}^n v_{jb}\Big)\Big).
\]

\section{Landau singularities}\label{Landau_app}

The Landau singularities of the simplex integral are best studied in the Lee-Pomeransky representation \eqref{LPrep}.  
They follow from solving simultaneously for all $v_{ij}$ the conditions
\[
0=\mathcal{U}+\mathcal{F},\qquad 0 = v_{ij}\frac{\partial}{\partial v_{ij}}(\mathcal{U}+\mathcal{F}).
\]
Here, the first Landau equation stipulates the vanishing of the Lee-Pomeransky denominator, while the second requires that this vanishing is either a double zero (for $v_{ij}\neq 0$), corresponding to a pinching of the $v_{ij}$ integration contour between two converging singularities of the integrand, or else a pinch of the integration contour between a singularity and the end-point of the integration ($v_{ij}=0$).  The second condition thus ensures the singularity generated by the vanishing denominator cannot be avoided by a deformation of the integration contour. 
Where the Landau conditions have more than one solution, the 
 solution with the greatest number of $v_{ij}\neq 0$ is  referred to as the {\it leading} singularity.

An important feature of the $\mathcal{U}$ polynomial  \eqref{Kirchhoff1} is that it is {\it multilinear} in the $v_{kl}$: from  the determinant structure  one sees that all the quadratic $v_{kl}^2$ terms cancel, and that no higher powers can appear since $v_{kl}$ enters only in the row/columns $(k,k)$, $(k,l)$, $(l,k)$ and $(l,l)$. 
Alternatively, this result follows from the matrix tree theorem where the Kirchhoff polynomial  $\mathcal{U}$ is the generator of spanning trees on the simplex.
Since $\mathcal{U}$ is also homogeneous of degree $(n-1)$, it follows that  
\[\label{Uid}
\sum_{k<l}\frac{\partial\mathcal{U}}{\partial v_{kl}}v_{kl} = (n-1)\mathcal{U}.
\]
We now find 
\[\label{Landau1eval}
\mathcal{U}+\mathcal{F} = \mathcal{U}+\sum_{k<l}\frac{\partial\mathcal{U}}{\partial v_{kl}}V_{kl} = \sum_{k<l}\frac{\partial\mathcal{U}}{\partial v_{kl}}\Big(\frac{v_{kl}}{n-1}+V_{kl}\Big)
\]
and so a solution of the first Landau condition for all $k<l$ is 
\[
v_{kl} = \lambda \,V_{kl} =- \lambda \,\bs{p}_k\cdot\bs{p}_l \qquad \mathrm{and}\qquad |G| = |\bs{p}_k\cdot\bs{p}_l|=0
\]
for some constant $\lambda$.  Evaluating 
the second Landau condition  on this solution $(\ast)$ of the first gives
\[
 \Big[v_{ij}\frac{\partial}{\partial v_{ij}}(\mathcal{U}+\mathcal{F})\Big]_{\ast}
=\Big[ v_{ij}\frac{\partial\mathcal{U}}{\partial v_{ij}}+v_{ij}\sum_{k<l}\frac{\partial^2\mathcal{U}}{\partial v_{ij}\partial v_{kl}}V_{kl}\Big]_{\ast}=
(\lambda +n-2)
\lambda^{n-2}V_{ij}\frac{\partial\mathcal{U}|_{v\rightarrow V
}}{\partial V_{ij}}
\]
using again the homogeneity of $\mathcal{U}$.  The second Landau condition is thus solved for all $i,j$ when $\lambda = 2-n$, and indeed this is the leading singularity since the $v_{kl}$ are generically nonzero.   Returning to \eqref{Landau1eval}, on the solution $(\ast)$ we have 
\[
(\mathcal{U}+\mathcal{F})_\ast= (2-n)^{n-2}|G|=0,
\] 
so to solve the first Landau condition we do indeed need the Gram determinant  to vanish.
Generally this requires analytic continuation to non-physical momentum configurations, since the only physical configurations (in Euclidean signature) for which the Gram determinant vanishes are collinear ones, and on physical grounds there are no collinear singularities.   There is no contradiction here since the Landau equations are necessary, but not sufficient, conditions for a singularity.

\section{Bernstein-Sato operators} 

\label{app:Bernstein}

In this appendix, we construct a Cayley-Menger analogue of the  classic  identity
\[\label{Cayley}
\mathrm{det(\partial)} (\mathrm{det}\,X)^a = a(a+1)\ldots (a+n-1) (\mathrm{det}X)^{a-1},
\]
where $X=(x_{ij})$ is an $n\times n$ matrix of independent variables and $\partial=(\partial/\partial x_{ij})$ is the corresponding matrix of partial derivatives.   For proofs and variants of this identity, traditionally attributed to Cayley, see,  {\it e.g.,} \cite{Caracciolo_2013, Fulmek}.  From a modern perspective,  \eqref{Cayley} is an example of a Bernstein-Sato operator, a  
differential operator whose action lowers the power $a$ to which some polynomial of interest is raised, generating in the process an auxiliary polynomial in $a$ known as the $b$-function \cite{Budur}.  Thus we have
\[
\mathcal{B}_f\, f(x_{ij})^a=b_f(a)f(x_{ij})^{a-1}
\] 
where for \eqref{Cayley}, $\mathcal{B}_f=\mathrm{det}(\partial)$, $f=\mathrm{det}\,X$ and $b_f(a) = a(a+1)\ldots (a+n-1)$.
In the following, we construct analogous operators for the Cayley-Menger determinant and other polynomials arising in our parametric representations \eqref{Cay2} and \eqref{Schwrep}.  Such relations are potentially a source of new weight-shifting operators, see {\it e.g.,} \cite{Tkachov:1996wh, Bitoun:2017nre}. 

Our starting point is the observation that
\[
\mathcal{B}_{|m|} = (|g|)\Big|_{v_{ij}\rightarrow \partial_{s_{ij}}}
\]
is a Bernstein-Sato operator for the Cayley-Menger determinant,
\[\label{Bernstein1}
\mathcal{B}_{|m|}\, |m|^a = b_{|m|}(a) |m|^{a-1}, \qquad b_{|m|}(a) =- \prod_{k=1}^{n-1}(1-k-2a).
\]
The operator $\mathcal{B}_{|m|}$ thus corresponds to evaluating the Kirchhoff polynomial $\mathcal{U}=|g|$ and replacing all $v_{ij}\rightarrow \partial_{s_{ij}}$ to generate a polynomial differential operator  in the $\partial_{s_{ij}}$.   We have verified  \eqref{Bernstein1} by direct calculation for matrices up to and including $n=5$.  Moreover, the leading behaviour at order $a^{n-1}$ follows by noting that such terms can only arise from all $n-1$ partial derivatives in $\mathcal{B}_{|m|}$ hitting a power of $|m|$ rather than a derivative of $|m|$.  Using \eqref{vdef} in the form  $\partial_{s_{ij}}|m|^a = a v_{ij} |m|^{a}$ along with \eqref{mginvreln}, then gives
\[\label{leadinga}
\mathcal{B}_{|m|}\,|m|^{a} = a^{n-1}|m|^{a}|g|+ O(a^{n-2})= (-1)^n \,2^{n-1}a^{n-1}|m|^{a-1}+ O(a^{n-2})
\]
in agreement with \eqref{Bernstein1}.\footnote{  
A full proof of \eqref{Bernstein1} 
likely follows via the methods of \cite{Caracciolo_2013},
though we will not pursue this here. }

Similarly, we find
\[
\mathcal{B}_{|g|} = (|m|)\Big|_{s_{ij}\rightarrow \partial_{v_{ij}}}
\]
({\it i.e.,} the Cayley-Menger determinant replacing each $s_{ij}\rightarrow \partial_{v_{ij}}$) 
is the Bernstein-Sato operator for the Kirchhoff polynomial $\mathcal{U}=|g|$,
\[\label{Bernstein2}
\mathcal{B}_{|g|} |g|^a = b_{|g|}(a) |g|^{a-1},\qquad b_{|g|}(a) =- \prod_{k=1}^{n-1}(1-k-2a).
\]
The $b$-function here is the same as that in \eqref{Bernstein1}, and the leading $a^{n-1}$ behaviour can be understood via the  analogous argument to that in \eqref{leadinga}.  
We note the result \eqref{Bernstein2} is equivalent to Theorem 2.15 of \cite{Caracciolo_2013}, since  
$|m| = |m^{(n+1,n+1)}|-|m^{(n+1,n+1)}+J|$ where $J$ is the $n\times n$ all-1s matrix, and $|m^{(n+1,n+1)}|$ is  the Cayley-Menger minor formed by deleting the final row and column consisting of 1s and 0s.
In addition, we find 
\[
\mathcal{B}_{|g|} (\partial_{v_{ij}}|g|)^a = 0.
\]

Some further results worth recording are the following. For the second minors of the Laplacian matrix, $|\tilde{g}^{(ij,ij)}|=\partial_{v_{ij}}|g|$, we find the operator
\[
\mathcal{B}_{\partial_{v_{ij}}|g|} = (\partial_{s_{ij}}|m|)\Big|_{s_{kl}\rightarrow \partial_{v_{kl}}}
\]
satisfies
\begin{align}\label{Bern3}
\mathcal{B}_{\partial_{v_{ij}}|g|}
\,(\partial_{v_{ij}}|g|)^a &= b(a)\, (\partial_{v_{ij}}|g|)^{a-1}, \qquad 
\mathcal{B}_{\partial_{v_{ij}}|g|} \,|g|^a =b(a)v_{ij} |g|^{a-1} 
\end{align}
where the $b$-function is proportional to that in \eqref{Bernstein1} but is missing the final factor, 
\[\label{bfn2}
b(a) = 2\prod_{k=1}^{n-2}(1-k-2a).
\]
Again, we have verified these identities for values up to and including $n=5$.  
This operator further annihilates all $(\partial_{v_{kl}}|g|)^a$
corresponding to other legs, {\it i.e.,} 
\[
\mathcal{B}_{\partial_{v_{ij}}|g|}(\partial_{v_{kl}}|g|)^a =0 \qquad \mathrm{for\,\, all}\quad (i,j)\neq (k,l).
\]
Similarly,
\[
(\partial_{v_{ij}}|g|)\Big|_{v_{kl}\rightarrow \partial_{s_{kl}}} |m|^a = b(a) s_{ij} |m|^{a-1}
\]
with the same $b$-function \eqref{bfn2}, 
but this operator does not appear to act simply (for $n>3$) on 
$(\partial_{s_{ij}}|m|)^a$, in contrast to \eqref{Bern3}.  Finding a Bernstein-Sato operator for  $(\partial_{s_{ij}}|m|)^a$ would be useful since by \eqref{dmdstominnors} this corresponds to the Cayley-Menger minors featuring in \eqref{Cay2}.

In principle, given a Bernstein-Sato relation such as \eqref{Bernstein1}, one might hope to apply it inside the parametric representation \eqref{Cay2} and integrate by parts to obtain an operator acting solely on the Schwinger exponential.  Since the exponential is diagonal in the representation \eqref{Cay2}, the result could then be translated to a differential operator in the external momenta.  This would then yield a new weight-shifting operator.

In practice, however, we must account for all the other powers of Cayley-Menger minors present in \eqref{Cay2}, as well as the arbitrary function.
Either we must find a modified Bernstein-Sato operator that acts appropriately on the {\it entire} non-exponential prefactor in \eqref{Cay2}, which seems hard to do, or else we must find some means of removing and then restoring these other factors.  The Cayley-Menger minors, for example, can be removed and then restored via a conjugation $\Omega\,\mathcal{B}_{|m|}\Omega^{-1}$ where $\Omega=\prod_{i<j}^n |m^{(i,j)}|^{-\alpha_{ij}-1}$.  After multiplying out, however, this conjugated operator is not in the Weyl algebra ({\it i.e.,} is non-polynomial in the $s_{ij}$ and their derivatives) and so does not trivially translate into an operator in the external momenta.  On the other hand, if we include additional powers of the $|m^{(i,j)}|$ on the left, so as to recover an operator in the Weyl algebra, besides lowering $\alpha$ in \eqref{Cay2} we also lower some of the $\alpha_{ij}$.  The operator then does not lower the spacetime dimension $d$.  
Thus we have not succeeded in finding new weight-shifting operators via this route, though with some variation the method might yet  be successful.

\bibliographystyle{JHEP}
\bibliography{CayleyPaper}

\providecommand{\href}[2]{#2}\begingroup\raggedright\begin{thebibliography}{10}

\bibitem{Antoniadis:2011ib}
I.~Antoniadis, P.~O. Mazur, and E.~Mottola, {\it {Conformal invariance, dark
  energy, and CMB non-Gaussianity}},  {\em JCAP} {\bf 09} (2012) 024,
  [\href{http://arxiv.org/abs/1103.4164}{{\tt arXiv:1103.4164}}].

\bibitem{Maldacena:2011nz}
J.~M. Maldacena and G.~L. Pimentel, {\it {On graviton non-Gaussianities during
  inflation}},  {\em JHEP} {\bf 09} (2011) 045,
  [\href{http://arxiv.org/abs/1104.2846}{{\tt arXiv:1104.2846}}].

\bibitem{Creminelli:2011mw}
P.~Creminelli, {\it {Conformal invariance of scalar perturbations in
  inflation}},  {\em Phys. Rev. D} {\bf 85} (2012) 041302,
  [\href{http://arxiv.org/abs/1108.0874}{{\tt arXiv:1108.0874}}].

\bibitem{Bzowski:2012ih}
A.~Bzowski, P.~McFadden, and K.~Skenderis, {\it {Holography for inflation using
  conformal perturbation theory}},  {\em JHEP} {\bf 04} (2013) 047,
  [\href{http://arxiv.org/abs/1211.4550}{{\tt arXiv:1211.4550}}].

\bibitem{Mata:2012bx}
I.~Mata, S.~Raju, and S.~Trivedi, {\it {CMB from CFT}},  {\em JHEP} {\bf 07}
  (2013) 015, [\href{http://arxiv.org/abs/1211.5482}{{\tt arXiv:1211.5482}}].

\bibitem{Kehagias:2012pd}
A.~Kehagias and A.~Riotto, {\it {Operator product expansion of inflationary
  correlators and conformal symmetry of de Sitter}},  {\em Nucl.Phys.} {\bf
  B864} (2012) 492--529, [\href{http://arxiv.org/abs/1205.1523}{{\tt
  arXiv:1205.1523}}].

\bibitem{McFadden:2013ria}
P.~McFadden, {\it {On the power spectrum of inflationary cosmologies dual to a
  deformed CFT}},  {\em JHEP} {\bf 10} (2013) 071,
  [\href{http://arxiv.org/abs/1308.0331}{{\tt arXiv:1308.0331}}].

\bibitem{Ghosh:2014kba}
A.~Ghosh, N.~Kundu, S.~Raju, and S.~P. Trivedi, {\it {Conformal Invariance and
  the Four Point Scalar Correlator in Slow-Roll Inflation}},  {\em JHEP} {\bf
  07} (2014) 011, [\href{http://arxiv.org/abs/1401.1426}{{\tt
  arXiv:1401.1426}}].

\bibitem{Anninos:2014lwa}
D.~Anninos, T.~Anous, D.~Z. Freedman, and G.~Konstantinidis, {\it {Late-time
  Structure of the Bunch-Davies De Sitter Wavefunction}},  {\em JCAP} {\bf
  1511} (2015), no.~11 048, [\href{http://arxiv.org/abs/1406.5490}{{\tt
  arXiv:1406.5490}}].

\bibitem{Arkani-Hamed:2015bza}
N.~Arkani-Hamed and J.~Maldacena, {\it {Cosmological Collider Physics}},
  \href{http://arxiv.org/abs/1503.08043}{{\tt arXiv:1503.08043}}.

\bibitem{Arkani-Hamed:2018kmz}
N.~Arkani-Hamed, D.~Baumann, H.~Lee, and G.~L. Pimentel, {\it {The Cosmological
  Bootstrap: Inflationary Correlators from Symmetries and Singularities}},
  \href{http://arxiv.org/abs/1811.00024}{{\tt arXiv:1811.00024}}.

\bibitem{Baumann:2019oyu}
D.~Baumann, C.~Duaso~Pueyo, A.~Joyce, H.~Lee, and G.~L. Pimentel, {\it {The
  Cosmological Bootstrap: Weight-Shifting Operators and Scalar Seeds}},
  \href{http://arxiv.org/abs/1910.14051}{{\tt arXiv:1910.14051}}.

\bibitem{Baumann:2020dch}
D.~Baumann, C.~Duaso~Pueyo, A.~Joyce, H.~Lee, and G.~L. Pimentel, {\it {The
  Cosmological Bootstrap: Spinning Correlators from Symmetries and
  Factorization}},  \href{http://arxiv.org/abs/2005.04234}{{\tt
  arXiv:2005.04234}}.

\bibitem{Sleight:2019hfp}
C.~Sleight and M.~Taronna, {\it {Bootstrapping Inflationary Correlators in
  Mellin Space}},  \href{http://arxiv.org/abs/1907.01143}{{\tt
  arXiv:1907.01143}}.

\bibitem{Raju:2012zr}
S.~Raju, {\it {New Recursion Relations and a Flat Space Limit for AdS/CFT
  Correlators}},  {\em Phys. Rev. D} {\bf 85} (2012) 126009,
  [\href{http://arxiv.org/abs/1201.6449}{{\tt arXiv:1201.6449}}].

\bibitem{Farrow:2018yni}
J.~A. Farrow, A.~E. Lipstein, and P.~McFadden, {\it {Double copy structure of
  CFT correlators}},  {\em JHEP} {\bf 02} (2019) 130,
  [\href{http://arxiv.org/abs/1812.11129}{{\tt arXiv:1812.11129}}].

\bibitem{Lipstein:2019mpu}
A.~E. Lipstein and P.~McFadden, {\it {Double copy structure and the flat space
  limit of conformal correlators in even dimensions}},  {\em Phys. Rev. D} {\bf
  101} (2020), no.~12 125006, [\href{http://arxiv.org/abs/1912.10046}{{\tt
  arXiv:1912.10046}}].

\bibitem{Armstrong:2020woi}
C.~Armstrong, A.~E. Lipstein, and J.~Mei, {\it {Color/kinematics duality in
  AdS$_{4}$}},  {\em JHEP} {\bf 02} (2021) 194,
  [\href{http://arxiv.org/abs/2012.02059}{{\tt arXiv:2012.02059}}].

\bibitem{Albayrak:2020fyp}
S.~Albayrak, S.~Kharel, and D.~Meltzer, {\it {On duality of color and
  kinematics in (A)dS momentum space}},  {\em JHEP} {\bf 03} (2021) 249,
  [\href{http://arxiv.org/abs/2012.10460}{{\tt arXiv:2012.10460}}].

\bibitem{Bzowski:2015pba}
A.~Bzowski, P.~McFadden, and K.~Skenderis, {\it {Scalar 3-point functions in
  CFT: renormalisation, beta functions and anomalies}},  {\em JHEP} {\bf 03}
  (2016) 066, [\href{http://arxiv.org/abs/1510.08442}{{\tt arXiv:1510.08442}}].

\bibitem{Bzowski:2017poo}
A.~Bzowski, P.~McFadden, and K.~Skenderis, {\it {Renormalised 3-point functions
  of stress tensors and conserved currents in CFT}},  {\em JHEP} {\bf 11}
  (2018) 153, [\href{http://arxiv.org/abs/1711.09105}{{\tt arXiv:1711.09105}}].

\bibitem{Bzowski:2018fql}
A.~Bzowski, P.~McFadden, and K.~Skenderis, {\it {Renormalised CFT 3-point
  functions of scalars, currents and stress tensors}},  {\em JHEP} {\bf 11}
  (2018) 159, [\href{http://arxiv.org/abs/1805.12100}{{\tt arXiv:1805.12100}}].

\bibitem{Bzowski:2022rlz}
A.~Bzowski, P.~McFadden, and K.~Skenderis, {\it {A handbook of holographic
  4-point functions}},  \href{http://arxiv.org/abs/2207.02872}{{\tt
  arXiv:2207.02872}}.

\bibitem{Gillioz:2018mto}
M.~Gillioz, {\it {Momentum-space conformal blocks on the light cone}},  {\em
  JHEP} {\bf 10} (2018) 125, [\href{http://arxiv.org/abs/1807.07003}{{\tt
  arXiv:1807.07003}}].

\bibitem{Gillioz:2019iye}
M.~Gillioz, X.~Lu, M.~A. Luty, and G.~Mikaberidze, {\it {Convergent
  Momentum-Space OPE and Bootstrap Equations in Conformal Field Theory}},  {\em
  JHEP} {\bf 03} (2020) 102, [\href{http://arxiv.org/abs/1912.05550}{{\tt
  arXiv:1912.05550}}].

\bibitem{Gillioz:2020wgw}
M.~Gillioz, {\it {Conformal partial waves in momentum space}},  {\em SciPost
  Phys.} {\bf 10} (2021), no.~4 081,
  [\href{http://arxiv.org/abs/2012.09825}{{\tt arXiv:2012.09825}}].

\bibitem{Polyakov:1970xd}
A.~M. Polyakov, {\it {Conformal symmetry of critical fluctuations}},  {\em JETP
  Lett.} {\bf 12} (1970) 381--383.

\bibitem{Bzowski:2019kwd}
A.~Bzowski, P.~McFadden, and K.~Skenderis, {\it {Conformal $n$-point functions
  in momentum space}},  {\em Phys. Rev. Lett.} {\bf 124} (2020), no.~13 131602,
  [\href{http://arxiv.org/abs/1910.10162}{{\tt arXiv:1910.10162}}].

\bibitem{Bzowski:2020kfw}
A.~Bzowski, P.~McFadden, and K.~Skenderis, {\it {Conformal correlators as
  simplex integrals in momentum space}},  {\em JHEP} {\bf 01} (2021) 192,
  [\href{http://arxiv.org/abs/2008.07543}{{\tt arXiv:2008.07543}}].

\bibitem{Karateev:2017jgd}
D.~Karateev, P.~Kravchuk, and D.~Simmons-Duffin, {\it {Weight Shifting
  Operators and Conformal Blocks}},  {\em JHEP} {\bf 02} (2018) 081,
  [\href{http://arxiv.org/abs/1706.07813}{{\tt arXiv:1706.07813}}].

\bibitem{Bzowski:2015yxv}
A.~Bzowski, P.~McFadden, and K.~Skenderis, {\it {Evaluation of conformal
  integrals}},  {\em JHEP} {\bf 02} (2016) 068,
  [\href{http://arxiv.org/abs/1511.02357}{{\tt arXiv:1511.02357}}].

\bibitem{Bzowski:2013sza}
A.~Bzowski, P.~McFadden, and K.~Skenderis, {\it {Implications of conformal
  invariance in momentum space}},  {\em JHEP} {\bf 03} (2014) 111,
  [\href{http://arxiv.org/abs/1304.7760}{{\tt arXiv:1304.7760}}].

\bibitem{Dolan:2011dv}
F.~A. Dolan and H.~Osborn, {\it {Conformal Partial Waves: Further Mathematical
  Results}},  \href{http://arxiv.org/abs/1108.6194}{{\tt arXiv:1108.6194}}.

\bibitem{Weinzierl:2022eaz}
S.~Weinzierl, {\it {Feynman Integrals}},
  \href{http://arxiv.org/abs/2201.03593}{{\tt arXiv:2201.03593}}.

\bibitem{Lee:2013hzt}
R.~N. Lee and A.~A. Pomeransky, {\it {Critical points and number of master
  integrals}},  {\em JHEP} {\bf 11} (2013) 165,
  [\href{http://arxiv.org/abs/1308.6676}{{\tt arXiv:1308.6676}}].

\bibitem{kirchhoff1847}
G.~Kirchhoff, {\it {\"U}ber die aufl{\"o}sung der gleichungen, auf welche man
  bei der untersuchung der linearen vertheilung galvanischer str{\"o}me
  gef{\"u}hrt wird},  {\em Annalen der Physik} {\bf 148} (1847), no.~12
  497--508.

\bibitem{kirchhoff1958}
G.~Kirchhoff, {\it On the solution of the equations obtained from the
  investigation of the linear distribution of galvanic currents},  {\em IRE
  transactions on circuit theory} {\bf 5} (1958), no.~1 4--7.

\bibitem{fiedler_2011}
M.~Fiedler, {\em Simplex geometry}.
\newblock Encyclopedia of Mathematics and its Applications.
\newblock Cambridge University Press, 2011.

\bibitem{Devriendt_2022}
K.~Devriendt, {\it Effective resistance is more than distance: Laplacians,
  simplices and the schur complement},  {\em Linear Algebra and its
  Applications} {\bf 639} (2022) 24--49.

\bibitem{dorfler2012kron}
F.~Dorfler and F.~Bullo, {\it Kron reduction of graphs with applications to
  electrical networks},  {\em IEEE Transactions on Circuits and Systems I:
  Regular Papers} {\bf 60} (2012), no.~1 150--163,
  [\href{http://arxiv.org/abs/1102.2950}{{\tt 1102.2950}}].

\bibitem{Coriano:2019sth}
C.~Corian\`o and M.~M. Maglio, {\it {On Some Hypergeometric Solutions of the
  Conformal Ward Identities of Scalar 4-point Functions in Momentum Space}},
  {\em JHEP} {\bf 09} (2019) 107, [\href{http://arxiv.org/abs/1903.05047}{{\tt
  arXiv:1903.05047}}].

\bibitem{Loebbert:2020hxk}
F.~Loebbert, J.~Miczajka, D.~M\"uller, and H.~M\"unkler, {\it {Massive
  Conformal Symmetry and Integrability for Feynman Integrals}},  {\em Phys.
  Rev. Lett.} {\bf 125} (2020), no.~9 091602,
  [\href{http://arxiv.org/abs/2005.01735}{{\tt arXiv:2005.01735}}].

\bibitem{Rigatos:2022eos}
K.~C. Rigatos and X.~Zhou, {\it {Yangian Symmetry in Holographic Correlators}},
   {\em Phys. Rev. Lett.} {\bf 129} (2022), no.~10 101601,
  [\href{http://arxiv.org/abs/2206.07924}{{\tt arXiv:2206.07924}}].

\bibitem{Itzykson:1980rh}
C.~Itzykson and J.~B. Zuber, {\em {Quantum Field Theory, pp 294-7}}.
\newblock International Series In Pure and Applied Physics. McGraw-Hill, New
  York, 1980.

\bibitem{Caracciolo_2013}
S.~Caracciolo, A.~D. Sokal, and A.~Sportiello, {\it Algebraic/combinatorial
  proofs of cayley-type identities for derivatives of determinants and
  pfaffians},  {\em Advances in Applied Mathematics} {\bf 50} (2013), no.~4
  474--594.

\bibitem{Fulmek}
M.~Fulmek, {\it {A combinatorial proof for Cayley's identity}},  {\em
  {Electronic Journal of Combinatorics}} {\bf 21} (2014), no.~4
  [\href{http://arxiv.org/abs/1309.6801}{{\tt arXiv:1309.6801}}].

\bibitem{Budur}
N.~Budur, {\em {Bernstein-Sato polynomials (Lecture Notes UPC, Barcelona)
  2015}}.
\newblock
  {\url{https://perswww.kuleuven.be/~u0089821/Barcelona/BarcelonaNotes.pdf}}.

\bibitem{Tkachov:1996wh}
F.~V. Tkachov, {\it {Algebraic algorithms for multiloop calculations. The First
  15 years. What's next?}},  {\em Nucl. Instrum. Meth. A} {\bf 389} (1997)
  309--313, [\href{http://arxiv.org/abs/hep-ph/9609429}{{\tt hep-ph/9609429}}].

\bibitem{Bitoun:2017nre}
T.~Bitoun, C.~Bogner, R.~P. Klausen, and E.~Panzer, {\it {Feynman integral
  relations from parametric annihilators}},  {\em Lett. Math. Phys.} {\bf 109}
  (2019), no.~3 497--564, [\href{http://arxiv.org/abs/1712.09215}{{\tt
  arXiv:1712.09215}}].

\end{thebibliography}\endgroup

\end{document}